\pdfoutput=1
\documentclass[12pt,  letterpaper]{article}

\usepackage{amsfonts}
\usepackage{subfigure,epsfig,amsthm,amsmath,latexsym,amssymb,url}
\usepackage{verbatim,setspace,multirow,fancyhdr,xfrac}%
\usepackage{pdfsync,epstopdf}
\usepackage{fullpage}
\usepackage{rotating}
\usepackage[round,comma]{natbib}

\newcommand{\noi}{\noindent}

\newcounter{Lcount}

\title{Combining Outcome-Based and Preference-Based Matching: A Constrained Priority Mechanism \\ \vspace{0.5cm}
\normalsize 
Forthcoming, \emph{Political Analysis} \\ \vspace{0.75cm}}

\author{Avidit Acharya\thanks{Associate Professor, Department of Political Science, 616 Serra Street Encina Hall West, Room 100, Stanford, CA 94305, USA. Email: \textsf{avidit@stanford.edu}} \and Kirk Bansak\thanks{Assistant Professor, Department of Political Science, University of California San Diego, 9500 Gilman Drive, La Jolla, CA 92093, USA. Email: \textsf{kbansak@ucsd.edu}} \and Jens Hainmueller\thanks{Professor, Department of Political Science, 616 Serra Street Encina Hall West, Room 100, Stanford, CA 94305, USA. Email: \textsf{jhain@stanford.edu}}}

\date{\today}

\begin{document}

\maketitle
\thispagestyle{empty}
\setcounter{page}{0}

\begin{abstract}
We introduce a constrained priority mechanism that combines outcome-based matching from machine-learning with preference-based allocation schemes common in market design. Using real-world data, we illustrate how our mechanism could be applied to the assignment of refugee families to host country locations, and kindergarteners to schools. Our mechanism allows a planner to first specify a threshold $\bar g$ for the minimum acceptable average outcome score that should be achieved by the assignment. In the refugee matching context, this score corresponds to the predicted probability of employment, while in the student assignment context it corresponds to standardized test scores. The mechanism is a priority mechanism that considers both outcomes and preferences by assigning agents (refugee families, students) based on their preferences, but subject to meeting the planner's specified threshold. The mechanism is both strategy-proof and constrained efficient in that it always generates a matching that is not Pareto dominated by any other matching that respects the planner's threshold. 
\end{abstract}

\clearpage

\section{Introduction}

We introduce a priority mechanism that matches agents to locations in instances where a planner/designer (hereafter, planner) can set a minimum acceptable threshold on her own measure of aggregate welfare. The design of our mechanism is motivated by the assignment of refugee families to host country locations. In this context, refugee families have preferences over locations, and host governments would like to conduct the assignment to take account of these preferences, but these governments would also like to make sure that their own measure of social welfare is not compromised so much so that it falls below a pre-specified threshold. In the refugee assignment problem, host country governments may consider their measure of social welfare to be an index of predicted integration success as measured by, for example, employment or earnings. In other applications such as student assignment to schools, this measure of welfare could be the average GPA of students, or their performance in standardized tests---measures that are typically of concern to school boards. 

Our mechanism is a priority mechanism but differs from the canonical version \citep[e.g.][]{satterthwaite1981strategy}  in the following respects. After preferences are elicited from the agents and the agents are lined up in a random order, each successive agent is assigned to their highest ranked location provided that assigning them to that location meets two conditions: (i) there is an available seat at that location, and (ii) there is a way to complete the assignment of the remaining agents that respects the planner's threshold. We assume that agents can rank locations strictly, except possibly their worst ranked locations. If there is no location that an agent can rank strictly that meets the two criteria above, then the agent is put in a ``holding set'' and will be assigned to one of their worst ranked locations (over which they are indifferent) at the end of the process. At this point, all agents in the holding set are assigned to locations to maximize the planner's welfare measure, and the assignment is complete.

Outcome-based matching was introduced in the context of refugee assignment to host country resettlement locations by \citet{bansak2018improving}.\footnote{Follow-up studies include \cite{trapp2018placement}, \cite{golz2019migration}, and \cite{bansak2020minimumrisk}.} The idea in outcome-based matching is to assign agents to locations so as to maximize a social planner's welfare measure, for example the refugee's expected employment success. Data-driven algorithms train supervised learners on historical data to discover synergies between places and types of refugees. The learned models are then used for newly arriving refugees to predict their expected integration success and optimally match them to locations where they have the highest probability of success subject to capacity and other constraints. Outcome-based matching is appealing because it harnesses historical data to maximize expected integration success and does not require collecting data on refugee preferences. Indeed, the outcome-based refugee matching methods as proposed by \citet{bansak2018improving} have already been implemented in the real world by research teams in collaboration with resettlement organizations. One implementation was conducted by the Swiss State Secretariat of Migration in collaboration with the \cite{bansak2018improving} research team. Another implementation of the methods proposed by \citet{bansak2018improving} was conducted by \cite{trapp2018placement} with HIAS, a resettlement agency in the United States. However, a pure outcome-based approach does not take preferences into account and does not utilize private information that refugees may possess regarding which location would work best for them. 

Our mechanism addresses this limitation by assigning agents based on their preferences, to an extent that is acceptable to the planner.  It draws on the strengths of both the pure preference-based approach and the data-driven outcome-based approach, allowing the planner to harness the power of data-driven assignment to ensure some minimum level of welfare while taking into account the preferences of the agents. The mechanism achieves this by integrating the data-driven matching algorithm of \cite{bansak2018improving} into a priority mechanism for preference-based matching. 

Our mechanism has several desirable properties. First, it strikes a compromise between the need of the planner to ensure a minimum level for their measure of average welfare, and the appeal of incorporating agents' preferences.\footnote{The idea of integrating machine learning methods with the preference-based matching methods of market design has been suggested by \cite{milgrom2018artificial}.}  Second, despite the added complexity of accounting for the planner's constraint, our mechanism inherits the desirable properties of priority mechanisms. It remains strategy-proof and hence is immune to strategic manipulation through false reporting of preferences. It is constrained Pareto-efficient  in that it generates an assignment that is not Pareto dominated by another assignment that also satisfies the planner's constraint.  It also allows agents to express preferences without the requirement that they strictly rank all locations. This flexibility is important, especially in the refugee assignment context, since there may be a large degree of heterogeneity as to whether refugees have distinct preferences over all locations. Finally, the mechanism is both computationally and administratively feasible.  It can be implemented by the planner with only minor adjustments to existing methods. It only requires the additional step of eliciting agents' strict preferences.

We provide two applications of our mechanism using data from two distinct settings. In the first, we illustrate how our mechanism could be used to assign refugees admitted into the United States to American cities, taking the planner's welfare measure to the be expected level of employment of a member of the refugee household within 90 days of resettlement. The importance of matching refugees to host country locations as a tool to improve integration success is discussed in \cite{mousa2018boosting}, and there have been many proposals for how host countries may approach the matching problem \citep[e.g.][]{moraga2014tradable,fernandez2015tradable,delacretaz2016refugee,andersson2016assigning,bansak2018improving,roth2018marketplaces,trapp2018placement,golz2019migration,bansak2020minimumrisk}. The idea of refugee matching is to select locations that are likely to be a good fit for a given refugee to thrive, and extant research has shown that the place of initial settlement has a profound impact on the long-term integration success of refugees \citep{aaslund2007and,damm2014neighborhood,bansak2018improving,marten2019ethnic}. 

In practice, however, the assignment of refugees in most countries is usually  determined by simple capacity constraints and/or proportional distribution keys. Governments want to ensure that refugees become self-sufficient and are typically reluctant to let them freely choose where to settle due to concerns that this could result in a highly uneven regional distribution and the creation of ethnic enclaves. That said, a few governments have started to appreciate the value of eliciting the refugee families' own preferences over locations.\footnote{For example, refugee families may possess valuable private information about which location would be best for them.}  Recognizing this value, the Dutch government, for example, has started collecting unstructured information on the location preferences of refugee families as part of their interviews. However, there currently exists no systematic data on refugee preferences, including in the United States. As a result, for our evaluation, we impute refugee preferences based on secondary migration data. 

In our second application, we demonstrate how our mechanism could be applied outside of refugee matching. In this application, we apply the mechanism to the problem of matching kindergarteners to schools in Tennessee, taking the planner's welfare measure to be the sum of their reading, math, and listening scaled scores on the Stanford Achievement Tests (SAT) for the Kindergarten level. School choice is a canonical application in the matching literature \citep[see, e.g.,][]{abdulkadirouglu2003school,abdulkadirouglu2009strategy, abdulkadiroglu2013matching,pathak2011mechanism,pathak2016really,ehlers2014school} and thus serves as a useful second context in which to illustrate our mechanism. 

Our paper contributes to the recent market design literature that takes into account a planner's constraints \citep{echenique2015control,kamada2015efficient,dur2018reserve}. Two papers along these lines are particularly related. The first, \cite{narita2019experiment}, looks at the problem of assigning subjects to treatments in a randomized control trial to maximize the welfare of the subjects subject to the constraint that the researcher glean a certain level of scientific information from running the trial. The second, \cite{delacretaz2016refugee}, considers several variants of the top trading cycles mechanism, first allowing for multidimensional constraints, and then allowing for the agents to have a starting endowment. Since these mechanisms do not respect both strategy-proofness and Pareto efficiency, they relax the efficiency requirement and impose the condition that there be no sequence of swaps that generate a Pareto improving assignment. Our paper differs from this prior work in that we incorporate preferences into the assignment problem while fixing a minimum expected outcome threshold.

Although our mechanism is both constrained efficient and strategy-proof, we also investigate how well we can do on a second metric of welfare, namely the percent of agents that receive one of their highly (e.g.~top-3) ranked locations. When we take a sample of re-randomizations of the priority order of agents, we find that there may be substantial potential gains to be made on this welfare metric.\footnote{We note, however, that since our constrained priority mechanism does not characterize the set of constrained efficient assignments (i.e.~the finding of \cite{abdulkadiroglu1998random} that every efficient assignment can be generated by some ordering of the agents does not generalize) this sampling approach may not give us an unbiased estimate of how much we can gain on this metric if we consider the set of all constrained efficient assignments.} We then suggest two ways of potentially capturing these gains without violating the requirement that the mechanism be strategy proof.\footnote{Note that a mechanism that re-randomizes the order of agents so that their order is decided based on the preferences that were elicited is not strategy-proof.} The first is to use historical data to predict the preferences of the agents based on their observable traits; then fix the ordering that does best on this metric under the predicted preferences; and, finally, elicit the agents' actual preferences and assign them according to this ordering. The second is to fix the ordering of agents so that an agent with a lower variance in outcome-scores across locations is served before one with a higher variance.

A final contribution of our study is to open a closer dialogue between political methodology and the study of market design. Political methodology has historically thrived on interdisciplinary engagement with methods developed in related disciplines, such as statistics, econometrics, psychometrics, and computer science. Yet, for some reason, it has largely neglected to engage with the foundational work that has developed in economics on the study of market and mechanism design.\footnote{A search on the \emph{Political Analysis} archive reveals zero search results for the terms ``market design'' or ``mechanism design''.} This is an unfortunate omission because market and mechanism design is arguably at the core of many issues that are highly relevant to political science. Fundamentally, market design is about engineering institutions to ensure that they generate desired outcomes, such as an efficient or equitable distribution of opportunities or resources. As economist Alvin Roth recently put it: market design is about ``Who gets what---and why" \citep{roth2015gets}. This phrase resembles one of the canonical definitions of politics as ``Who gets what, when, and how'' by Harold Lasswell \citep{laswell1936politics}. Institutional mechanisms that allocate opportunities and resources are a central feature of modern democracies, and algorithms are increasingly used for public policy in a wide variety of domains. We hope that our study can help pave a path for political methodology to begin to contribute to these important developments given its unique blend of expertise.

\section{The Mechanism}

\subsection{Preliminaries}

There are $n$ agents (refugee families/school children) randomly labeled $1,...,n$, each of which has to be assigned to a location (host country city or town/school). Let $L$ denote the finite set of locations. Each location $l \in L$ has a capacity $q_l \geq 1$ as to how many agents it can accommodate. We assume that $n \leq \sum_l q_l$ so that it is feasible to assign all agents. For each agent $i$, let $g_i(l)$ be a measure of success at location $l$ (employment probability/test scores) when assigned to that location. In practice, this measure may need to be estimated, in which case it represents an agent's success at location $l$ in expectation. This measure may be accounted for in the agent's preferences, but is the key consideration for a social planner. We refer to $g_i(l)$ as the planner's outcome score for agent $i$ at location $l$.

Each agent $i$ has a complete and transitive preference ordering $\succsim_i$ over the set of locations.\footnote{We assume that all agents prefer to be assigned to a location rather than be not assigned, so we can omit non-assignment from the set of possible outcomes for each agent.} Indifference and the strict preference relations are denoted $\sim_i$ and $\succ_i$, respectively, and $\succsim = (\succsim_1,...,\succsim_n)$ denotes the vector of preferences. 

We make the assumption on agents' preferences that the only indifferences are over the worst-ranked locations. That is, apart from possibly having ties among a set of locations that an agent deems to be the worst, each agent has a strict preference over all of the other locations. Formally, for all agents $i$, if $l \sim_i l'$ for some $l' \neq l$, there is no $l''$ such that $l \succ_i l''$. This still allows for an agent to be indifferent over all locations. This assumption is motivated by our application to refugee assignment: refugee families often do not have full information on all possible locations, but they may have (strict) preferences over a limited set of top choices.\footnote{We interpret this as reflecting true indifference across the worst-ranked locations. Our mechanism would not necessarily be strategy-proof if the agents do in fact have strict preferences over these locations but express indifference due to lack of information.} 

Define the set $S_i = L \backslash \{l \in L : \exists l' \sim_i l\}$ which are all of the locations except any that agent $i$ is indifferent over. Agent $i$ has a strict preference across all locations in $S_i$ and if any location is left out of $S_i$ then it must have been ranked worst.

A matching $\mu$ maps the set of agents to locations. A matching $\mu$ is 
\begin{enumerate}
\item \textbf{feasible} if it satisfies the capacity constraints: $|\mu^{-1} (l)| \leq q_l, \forall l$
\item $\bar g$\textbf{-acceptable} if the average outcome score is not lower than $\bar g$: $\frac{1}{n} \sum_i g_{i} (\mu(i)) \geq \bar g.$
\end{enumerate}
$\bar g$-acceptability reflects the idea that the planner wants the average outcome score not to fall below a specified threshold $\bar g$. The planner wants to ensure that the allocation is such that agents have some minimum level of expected outcomes (e.g. a minimum expected employment rate/GPA or test score). 

Note that not all values of $\bar g$ can produce a feasible matching. Let $\bar g^\text{max}$ denote the highest possible average outcome score that can be generated by a feasible matching:
\begin{equation} \label{eq:pmax}
\bar g^\text{max} := \max_{\mu} \frac{1}{n} \sum_i g_i ( \mu(i))  \text{ subject to } |\mu^{-1} (l)| \leq q_l, \forall l
\end{equation}
Feasible $\bar g$-acceptable matchings exist only for $\bar g \leq \bar g^\text{max}$.

\subsection{The Assignment Procedure}

Given a value of $\bar g \leq \bar g^{\max}$, the algorithm starts with agent 1 and works down to agent $n$ in a sequence of $n$ steps before completing in either the $n$th or an additional $(n+1)$th step. At  Step $i \leq n$, agent $i$ is either assigned to a location or put on hold by being added to a set of temporarily unassigned agents that will all get assigned simultaneously at Step $n+1$.  At each Step $i$, let $N_i$ denote the set of agents $j < i$ that have been put on hold. $N_1 = \emptyset$ since at the start of the algorithm no agent is on hold.

If agent $j < i$ was assigned a location prior to Step $i$, then let $\alpha_i (j)$ denote the location and $(j, \alpha_i(j))$ the assignment, viewing $\alpha_i$ as a function. Refer to this function as the completed assignment at Step $i$.  Note that $\alpha_1 = \emptyset$, so the completed assignment at Step 1 is trivial. A remaining assignment $\beta_i$ at Step $i$ is a mapping of the unassigned agents $\{i,...,n\} \cup N_i$ to locations such that
$$ \mu_{(\alpha_i, \beta_i)} (j) := \left\{\begin{array}{ll} \alpha_i(j) & \text{if $j < i$}\\
\beta_i(j) & \text{if $j \in \{i,...,n\} \cup N_i$} \end{array} \right. $$
is a matching. We refer to $\mu_{(\alpha_i, \beta_i)}$ as the matching associated with the pair of completed and remaining assignments $(\alpha_i, \beta_i)$. The existence of these matchings will be guaranteed recursively by the algorithm. 

At each Step $i \leq n$, given $\alpha_i$ define the set
\begin{align}
L^{\bar g}_i (\alpha_i) = \{  l \in  L : & \text{ $\exists$ $\beta_i$ s.t. $l = \beta_i (i)$  and $\mu_{(\alpha_i, \beta_i)}$ is} \text{ a feasible $\bar g$-acceptable matching} \} \nonumber
\end{align}
This is the set of locations that are not at full capacity and for which there is a way to finish assigning all unassigned agents so as to create a feasible $\bar g$-acceptable matching. 

Let $q^i_l$ be the remaining capacity of location $l$ after any agents ahead of $i$ (i.e., $j < i$) have been assigned in the previous $i-1$ steps. At the start we have $q^1_l = q_l$ for all $l$.  It will also be convenient to define the following problems: for all Steps $i=1,...,n+1$, and given a vector $q^i := (q^i_l)_{l \in L}$,
\begin{align} \label{eq:maxim}
G_i (q^i) := & \max_{\beta_i} \sum_{j \in \{i,...,n\} \cup N_i} g_j (\beta_i (j)) \text{ subject to } |\beta_i^{-1} (l) | \leq q^i_l,  \forall l 
\end{align}
with the convention that $\{i,...,n\} := \emptyset$ if $i = n+1$. At each Step $i$,  the problem in \eqref{eq:maxim} finds the remaining assignment that maximizes the total outcome score subject to the updated capacity constraints at Step $i$. The solution to this problem at each step determines whether the associated matching is $\bar g$-acceptable. In fact, to verify whether or not a location $l$ belongs in $L^{\bar g}_i(\alpha_i)$ we must first check whether the highest possible value of the average outcome score that can be achieved under the remaining assignment is at least $\bar g$; i.e., whether 
$$
\bar g_i(l) := \frac{1}{n} \left( G_{i+1}(q^{i+1}) + g_i (l) + \sum_{j < i \text{ s.t.~}j \notin  N_i} g_j (\alpha_i(j)) \right)  \geq \bar g
$$
where $q^{i+1}_{l'} = q^i_{l'}$ for all $l' \neq l$ and $q^{i+1}_{l}= q^i_{l} -1$.  
If indeed $\bar g_i(l) \geq \bar g$ and $q^i_{l} > 0$, then $l$ belongs to $L^{\bar g}_i (\alpha_i)$; otherwise it does not. Constructing $L^{\bar g}_i(\alpha_i)$ at each Step $i=1,...,n+1$ therefore requires solving the problems given in \eqref{eq:maxim}. In addition, to verify whether $\bar g < \bar g^{\max}$ also requires solving one of these problems since the problem in \eqref{eq:pmax} equals $G_1 (q^1)/n$.

The steps of the algorithm are as follows.\smallskip

\textbf{Step $0$.} Verify that $\bar g \leq \bar g^{\max}$ and proceed only if it holds. \smallskip

\textbf{Step $i \leq n$.} If $S_i \cap L^{\bar g}_i (\alpha_i)$ is empty (meaning that there is no location that agent $i$ ranked strictly to which it could be assigned, and we can find a remaining assignment that generates a feasible $\bar g$-acceptable matching), then place agent $i$ on hold. In this case, set 
\begin{center}
$N_{i+1} = N_i \cup \{i\}$, $\alpha_{i+1}= \alpha_{i}$, $q^{i+1}_l = q^i_l~\forall l$
\end{center}
and move on to Step $i+1$. Otherwise, if $S_i \cap L^{\bar g}_i (\alpha_i)$ is nonempty, then it contains a unique best location from the perspective of agent $i$ -- i.e., a location $l_i^*$ such that $l_i^* \succ_i l$ for all $l \in S_i \cap L^{\bar g}_i(\alpha_i)$. This follows from the fact that $i$ ranks the elements of $S_i$ strictly. Assign agent $i$ to $l_i^*$, and set 
\begin{align*}
\text{$N_{i+1} = N_i$},  \text{ $\alpha_{i+1}= \alpha_{i} \cup \{(i,l_i^*)\}$ },  \text{$q^{i+1}_{l_i^*} = q^i_{l_i^*}-1$, and } \text{$q^{i+1}_l = q^i_l$ } \text{$\forall l \neq l_i^*$}
\end{align*} 
If $i<n$, then move to Step $i+1$. If $i = n$, then move to Step $n+1$ only if a agent was ever put on hold (i.e., $N_{n+1} \neq \emptyset$); otherwise, stop. \smallskip

\textbf{Step $n+1$.} At this stage the only unassigned agents are those that were put on hold in $N_{n+1}$. Here, choose any remaining assignment that maximizes the average outcome score given the completed assignment and the capacity constraints; that is, solve \eqref{eq:maxim} for $i = n+1$ and stop.\medskip

For any preference vector $\succsim$ satisfying our assumptions, our algorithm produces a matching, namely $\mu_{(\alpha_s, \beta_s)}$, where $s \in \{n, n+1\}$ was the step at which the algorithm stopped. The algorithm defines a mechanism $\varphi$, which, given the other parameters of the model, is a mapping from preference vectors to feasible matchings. We refer to the mechanism as $\bar g$-constrained priority, since it is a modification of the usual priority mechanism \citep{satterthwaite1981strategy}.

At each Step $i$, implementation of the mechanism involves iteratively solving the maximization problem in Equation 2 to verify that $\bar g$-acceptability can still be met if agent $i$ were assigned to each available location in order of preference, until such a location is found. This amounts to iteratively solving a standard linear sum assignment problem, for which various polynomial-time algorithms exist.\footnote{In graph theory, the assignment problem is known as a maximum weighted bipartite matching. See the Supplemental Information (SI) for more details on how the assignment problem is featured in the mechanism implementation.} Under a worst-case scenario where every agent is put on hold after unsuccessfully considering all of its strictly ranked locations, this would require solving an equally sized maximization problem in Equation 2 a total of $n (|L|-2)$ times.\footnote{Note that the maximization problem would then need to be solved one final time at Step $n+1$ with all of the agents. The reason the worst-case scenario features the $(|L|-2)$ term is that it arises when agents have strictly ranked all but two locations, since it is not possible to strictly rank all but one location, and if all locations have been strictly ranked then agents will not be put on hold and the maximization problem in Equation 2 would gradually become smaller and less costly to solve at each successive Step $i$.}

\subsection{Properties of the Mechanism} \label{sec: properties}

Let $\varphi(\succsim)$ denote the matching produced by the $\bar g$-constrained priority mechanism for any preference vector $\succsim$ that satisfies our assumptions, and $\varphi(\succsim)(i)$ the location assignment of agent $i$ under this matching. By construction, the matching produced by this mechanism is feasible and $\bar g$-acceptable. In addition, the mechanism satisfies two key properties. It is:
\begin{enumerate}
\item \textbf{constrained efficient} in the sense that for all preference vectors $\succsim$ that satisfy our assumptions, $\varphi(\succsim)$ is not Pareto dominated by another feasible $\bar g$-acceptable matching $\mu$. That is, it is not the case that $\mu(i) \succsim_i \varphi(\succsim)(i)$ for all agents $i$, and $\mu (i) \succ_i \varphi(\succsim)(i)$ for some agent $i$.
\item \textbf{strategy-proof} in the sense that truthful reporting is a dominant strategy of the induced preference reporting game. That is, for every preference vector $\succsim$ satisfying our assumptions, every agent $i$, and every alternative preference $\succsim_i'$ that $i$ could report that also satisfies our assumptions, $\varphi(\succsim)(i) \succsim_i  \varphi(\succsim'_i, \succsim_{-i})(i)$.
\end{enumerate}

The proof that the mechanism is constrained efficient and strategy-proof is straightforward, but for completeness we include it in the SI.

One important property of the canonical priority mechanism that does \emph{not} carry over to our $\bar g$-constrained priority mechanism is the property that the mechanism characterizes the full set of Pareto efficient assignments. \cite{abdulkadiroglu1998random} showed that for any Pareto efficient assignment, there exists an ordering of agents under which implementing the priority mechanism for that ordering generates that assignment. Given this, one could ask whether for every $\bar g$-constrained efficient assignment, there exists an ordering of the agents for which the $\bar g$-constrained priority mechanism generates that assignment. The answer to this question turns out to be no, as demonstrated by the following example with two agents $1$ and $2$ and three locations $A, B$ and $C$. The table gives the ranking of the three locations for each agent and in parentheses the outcome score $g_i(l)$ for each agent-location pair. 

\begin{center}
\begin{tabular}{llll}
& 1st choice  & 2nd choice & 3rd choice \\
1 &   $A$ (0.1)  & $B$ (0.5) & $C$ (0.9)\\
2 & $A$ (0.1) & $C$ (0.5) & $B$ (0.9)
\end{tabular}
\end{center}
Suppose that each location has a capacity of 1 seat. If the planner's threshold $\bar g$ is set to $0.45$ and agent 1 goes first, then he will be assigned to location $A$, and agent 2 will be assigned to $B$. If agent 2 goes first then she will be assigned to $A$, and agent $1$ will be sent to $C$. But the possibility of sending 1 to $B$ and 2 to $C$ also meets the planner's constraint and is not Pareto dominated by any other assignment that is acceptable to the planner.

\section{Applications}

To illustrate the mechanism, we apply it both to simulated data as well as two empirical examples using real-world data from the United States that involve the assignment of refugees to resettlement locations and the assignment of students to schools.\footnote{Replication materials for this study are available in \cite{codeocean2020abh, DVN/ZEV0WX}.}

Our mechanism requires the planner to select a value for $\bar g$, and this choice implies a tradeoff between an outcome-based and preference-based matching. From the planner's perspective, it is desirable to achieve the highest possible value of $\bar g$ to ensure that the agents' outcomes are optimized. However, setting a higher value of $\bar g$ comes at the cost of assigning agents to locations that are, in expectation, lower in their preference rankings. That is, while the mechanism simultaneously considers both outcomes and preferences, there is a tradeoff between the two, where the balance of that tradeoff changes as $\bar g$ increases. 

The magnitude of the tradeoff also depends upon the joint distribution of agents' preference rankings and their outcome scores. Two measures, in particular, play an important role: the correlation between outcome scores and preference rankings within agents (i.e. the degree to which an agent's preferred locations align with the locations where that agent would achieve their best outcomes) and the correlation between preference rankings across agents (i.e. the degree to which agents have similar preference rankings). We demonstrate this below.

\subsection{Simulation Data}

We apply the mechanism to simulated data to show these properties. For simplicity, our simulations involve assigning 100 agents to 100 locations with one seat each. For each agent, we randomly generate a preference rank vector (with $1$ indicating the most desired location and $100$ the worst) and an outcome score vector (with values in $[0,1]$). The simulations vary both the correlation between preference and outcome vectors ($-0.5$, $0$, and $0.5$) and the correlation between preference vectors across agents ($0$, $0.5$, and $0.8$).\footnote{The correlation between preference and outcome vectors treats higher preferences (i.e. closer to $1$) as more positive values, such that a positive correlation between preferences and outcomes indicates more highly preferred locations are those that also result in higher outcome scores.} This yields nine different scenarios, and in each we apply our mechanism to make the assignment for various values of $\bar g$. See the Supplemental Information (SI) for details.

\subsection{Refugee Data}

As a simulated illustration of how the mechanism could perform in a real-world scenario, we apply it to data from refugees in the United States, where refugee families are the agents and resettlement cities are the locations. Early employment is a core goal of the U.S. resettlement program, which strives to quickly transition refugees into self-sufficiency after arrival. This application illustrates how our mechanism could hypothetically be employed in the United States to achieve a desired level of early employment while geographically assigning refugees based on their location preferences.

Our real-world refugee data includes de-identified information on working-age refugees (ages 18 to 64; N = 33,782) who have been resettled to the United States during the 2011-2016 period by one of the largest U.S. refugee resettlement agencies. Over this time period, the agencies' placement officers centrally assigned refugees to one of approximately 40 resettlement locations in the agency's network. The data contain details on the refugee characteristics such as age, gender, origin, and education. It also includes the assigned resettlement location, whether the refugee was employed at 90 days after arrival, and whether the refugee migrated from the initial location within 90 days. 

We applied our mechanism to data on the refugee families who arrived in the third quarter (Q3) of 2016, specifically focusing on refugees who were free to be assigned to different resettlement locations (561 families), in contrast to refugees who were predestined to specific locations on the basis of existing family or other ties. To generate each family's outcome score vector across each of the locations, we employed the same methodology in \cite{bansak2018improving}, using the data for the refugees who arrived from 2011 up to (but not including) 2016 Q3 to generate models that predict the expected employment success of a family (i.e. the mean probability of finding employment among working-age members of the family) at any of the locations, as a function of their background characteristics. These models were then applied to the families who arrived in 2016 Q3 to generate their predicted employment success at each location, which comprise their outcome score vectors. See the SI and \cite{bansak2018improving} for details.

Our mechanism also requires data on location preferences of refugees. To the best of our knowledge, such data do not currently exist in the United States, where refugees are assigned to locations by the resettlement agencies. We therefore infer revealed location preferences from secondary migration behavior. Specifically, we use the same modeling procedures used in the outcome score estimation, simply swapping in outmigration in place of employment as the response variable. This allows us to predict for each refugee family that arrived in 2016 Q3 the probability of outmigration at each location as a function of their background characteristics. For each family, we then rank locations such that the location with the lowest (highest) probability of outmigration is ranked first (last). Details about the data and sample are provided in the SI.

We acknowledge that inferred location preferences from secondary migration behavior are not necessarily the same as the stated location preferences that refugees would express in an application form if given the opportunity to do so by host country governments. That said, there are reasons to believe that the inferred location preferences provide a useful proxy. Outmigration is a costly signal indicating that a refugee prefers to move rather than stay in the originally assigned location. \cite{mossad2019search} provide a comprehensive analysis of the secondary migration patterns of refugees in the United States and find that secondary migration is mostly driven by refugees relocating in search of employment opportunities and co-ethnic communities. One of the main channels through which these effects operate is the refugee's nationality, which is also an important predictor in the model that we use to infer revealed location preferences from secondary migration.

\subsection{Education Data}

As a second simulated illustration of how the mechanism could perform in a real-world scenario, we apply it to education data from the United States, where the agents are individual students and the locations are schools. In particular, we consider data from the Tennessee's Student Teacher Achievement Ratio (STAR) project conducted by the Tennessee State Department of Education (for details see \citet{SIWH9F_2008}). These data contain information on the choice of elementary schools for a large sample of students as well as information on the test score performance of these students. We focus on the Kindergarten grade level and apply our mechanism to generate new assignments of students to schools with the goal of improving the outcomes of students as measured by standardized tests administered at the end of Kindergarten. One could imagine a school district setting a minimum test score that should be achieved on average.

To generate each student's outcome score vector across each of the schools, we employ the same methods as in the previous application to predict the expected test scores of a student at any of the schools, as a function of their background characteristics. The background characteristics included the students' age, gender, and race, as well as information on whether they are eligible for free school lunches (a proxy for socioeconomic status) or special education. The test score outcome was defined as the sum of reading, math, and listening scaled scores on the Stanford Achievement Tests (SAT) for the Kindergarten level.

We inferred revealed school preferences of students from the observed transfers out of the schools. Specifically, we used the same modeling procedure as for the test scores but instead used a response variable that measured whether a student had transferred to another school by the first, second, or third grade. Based on these models we can then predict for each student the propensity for leaving each school as a function of their background characteristics. For each student, we then rank schools such that the school for which they have the lowest (highest) propensity for transferring out is ranked first (last). 

We generate these outcome score and preference vectors and apply our mechanism to a random sample of 1,000
%for a total sample of 1,674 
students from 33 schools that are observed for all grades from Kindergarten through 3rd grade and have non-missing data for tests scores and background characteristics. Details about the data and sample are provided in the SI.

\section{Results}

\subsection{Simulated Data} \label{sec: simul}

\begin{figure}[t!]
\centering
\includegraphics[width=1\linewidth]{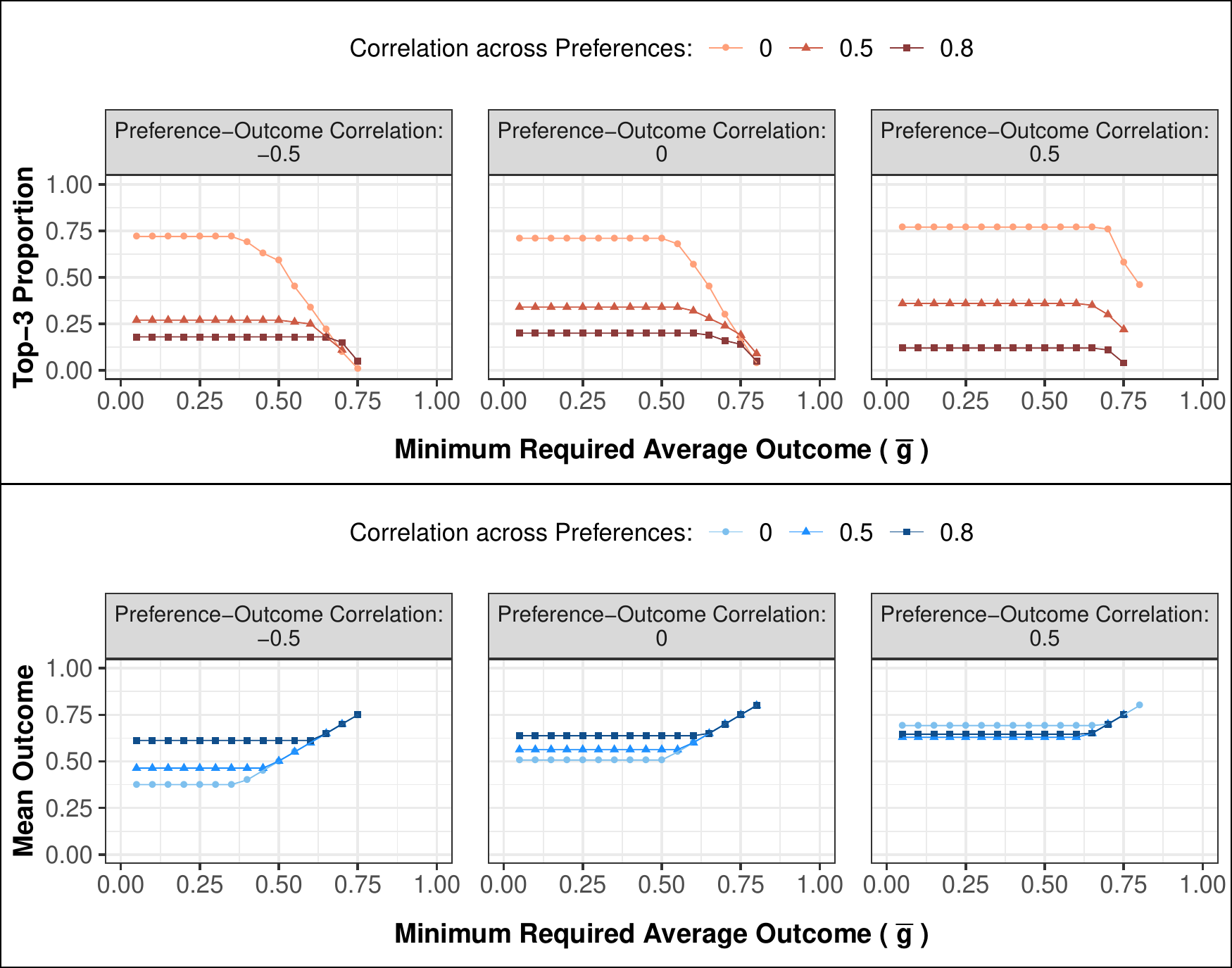}
\caption{\small Results from applying our $\bar g$-Constrained Priority Mechanism to simulated data that varies the correlations between preference and outcome score vectors and the correlations between preference vectors across agents. Upper panel shows the average probability that a agent was assigned to one of its top three locations. Lower panel shows the realized average outcome score. N=100.}
\label{fig:1}
\end{figure}

Figure \ref{fig:1} depicts the results for assignment under the $\bar g$-constrained priority mechanism for nine different simulation scenarios that vary the correlation between preferences and outcome scores within agents and the correlation between preferences across agents. In addition, to model a real-world scenario in which agents can indicate only a limited number of top locations in an application form, the preference vectors are truncated such that only the top 10 ranks are retained and indifference is established among the remaining location. The top panel shows the proportion of agents who were assigned to one of their top three locations given various levels of $\bar g$, the threshold for the minimum average outcome score. The bottom panel shows the mean outcome score for agents in their assigned locations for the same levels of $\bar g$. The curves end once $\bar g^{\max}$ has been reached and hence no feasible assignment is possible.

There is a clear tradeoff between realized preference ranks and outcome scores in all simulations. As $\bar g$ is increased, the realized mean outcome score eventually increases. This is a mechanical result of increasing $\bar g$ and hence enforcing the requirement for a higher mean outcome value. Simultaneously, as soon as the mean outcome score is impacted, the proportion of agents assigned to one of their preferred locations also begins to decrease. This occurs because enforcing the requirement for a higher value of $\bar g$ requires the mechanism to deviate from the preference-based optimization.

Figure \ref{fig:1} also shows how the immediacy and severity of the tradeoff can vary substantially depending upon the joint distribution of preferences and outcome scores.\footnote{It can also depend on the number of seats available in each location and the extent to which each location contributes to the correlations.} First, focusing on the top panel, we see that the higher the correlation between agents' preferences, the worse is the achievable baseline proportion of agents that can be assigned to one of their top locations at the lowest values of $\bar g$. This result, which holds regardless of the correlation between preferences and outcome scores, is intuitive: the more similar are different agents' preferences, the more rivalrous is the matching procedure, and hence the more difficult it is to match agents to one of their top-ranked locations given limited capacity in each location.

Second, the more positive the correlation between preferences and outcome scores, the less severe is the tradeoff in the sense that the tradeoff does not kick in until higher levels of $\bar g$ are enforced. The intuition for this result is that if preferences and outcomes are positively correlated, then matching based on preferences should indirectly also lead to outcome-based matching, and hence deviation from the preference-based solution will not occur until a higher level of $\bar g$ is reached. This is a useful finding from the standpoint of a real-world implementation of the mechanism. If, in advance of their preference reporting, agents were given information on their predicted outcomes in each location, they could incorporate such information into their preference determination. If this results in a closer alignment of preferences and outcomes, that would help alleviate the tradeoff in the mechanism.

Third, turning to the bottom panel in Figure \ref{fig:1}, we see that once the tradeoff kicks in, the realized mean outcome curves trace closely along the identity line; that is, upon enforcing a level of $\bar g$ that deviates from the preference-based assignment, the mechanism will find an alternative assignment that optimizes for preferences subject to just barely satisfying the $\bar g$ constraint. The realized mean outcome results also mirror the trends on realized preference ranks: The more positive is the correlation between preference and outcome vectors, the later the tradeoff kicks in.

Fourth, we see that given a negative correlation between preferences and outcome scores, the correlation across preference vectors has a significant impact on how the tradeoff affects the realized mean outcome score, with the tradeoff being more severe with a low correlation across preference vectors. This result can be explained as follows. A negative correlation between preference and outcome vectors implies that preference-based assignment is counter to the goal of optimizing for realized outcome scores. However, if there is also a positive correlation across agents' preferences, that means that different kinds of agents generally prefer the same locations, and hence also that the locations that result in low outcome scores are also similar across agents, thus limiting the degree to which matching based on preferences will actually hurt realized outcome scores on average. If, in contrast, there is no correlation across preferences, then there is greater latitude for the mechanism to assign agents to their higher-ranked locations, which also happen to be the locations that are the worst for their outcome scores. As the correlation between preference and outcome vectors becomes more positive, this dynamic begins to disappear. However, the reason it does not reverse in the bottom-right panel of Figure \ref{fig:1} is due to the existence of trailing indifferences in the preference rank vectors, which means the agents who could not be matched to one of their strictly ranked locations are assigned using outcome-based optimization, thereby limiting the effect of the phenomenon described above.\footnote{The SI includes the results of the same simulations without truncating the preference rank vectors. In that case, we do see the expected reversal across the lower three panels.}

\subsection{Refugee Data}

\begin{figure}[t!]
\centering
\includegraphics[width=0.85\linewidth]{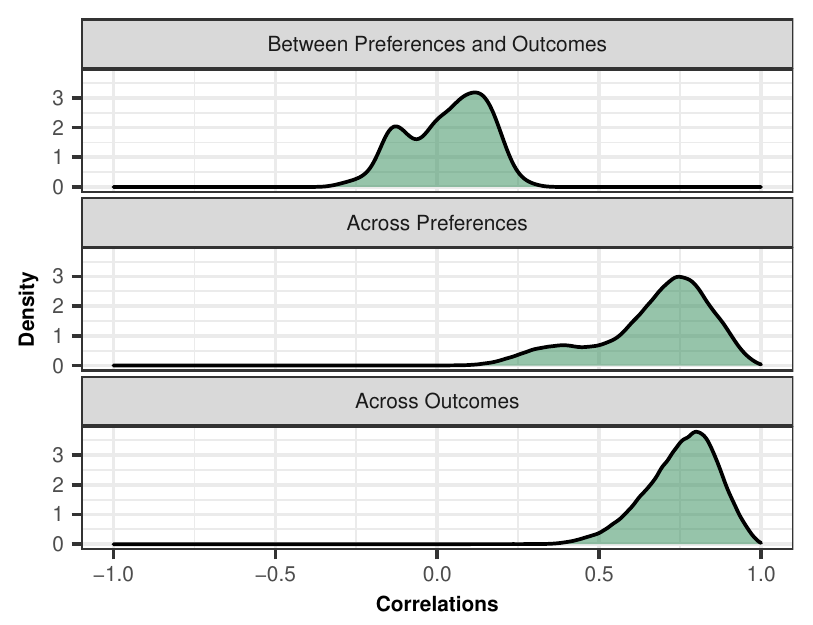}
\caption{\small Shows the distribution of pairwise correlations between refugee family location preferences, integration outcomes (i.e.~employment), and preferences and outcomes. N=561 refugee families who arrived in the United States in Q3 of 2016.}
\label{fig:2}
\end{figure}

Figure \ref{fig:2} shows features of the joint distribution of the refugee families' outcome score and preference rank vectors. The top panel pertains to the correlation between the families' outcome and preference vectors. For each family, a correlation is computed between its two vectors, and the panel displays the distribution of those correlations. The distribution is roughly centered around zero (the mean correlation is $0.03$). This suggests a relatively limited relationship between the locations refugees prefer and those where they would actually achieve better employment outcomes. This is an interesting finding and also has a key policy implication. Providing refugees with information on which locations are beneficial for their employment outcomes would allow them to formulate more informed preferences. If this results in a closer correlation between preference and outcome vectors, this would help strengthen our mechanism since a more positive correlation alleviates the tradeoff between outcome- and preference-based matching.

The middle panel in Figure \ref{fig:2} shows the distribution of pairwise correlations between families' preference vectors. The correlations are mostly highly positive, with a mean correlation of $0.67$. This shows that preference vectors are relatively similar across the families; many refugees would more or less prefer to be placed in similar locations. Given the existence of location capacity constraints, this is an inconvenient finding from the standpoint of preference-based assignment.

\begin{figure}[t!]
\centering
\includegraphics[width=1\linewidth]{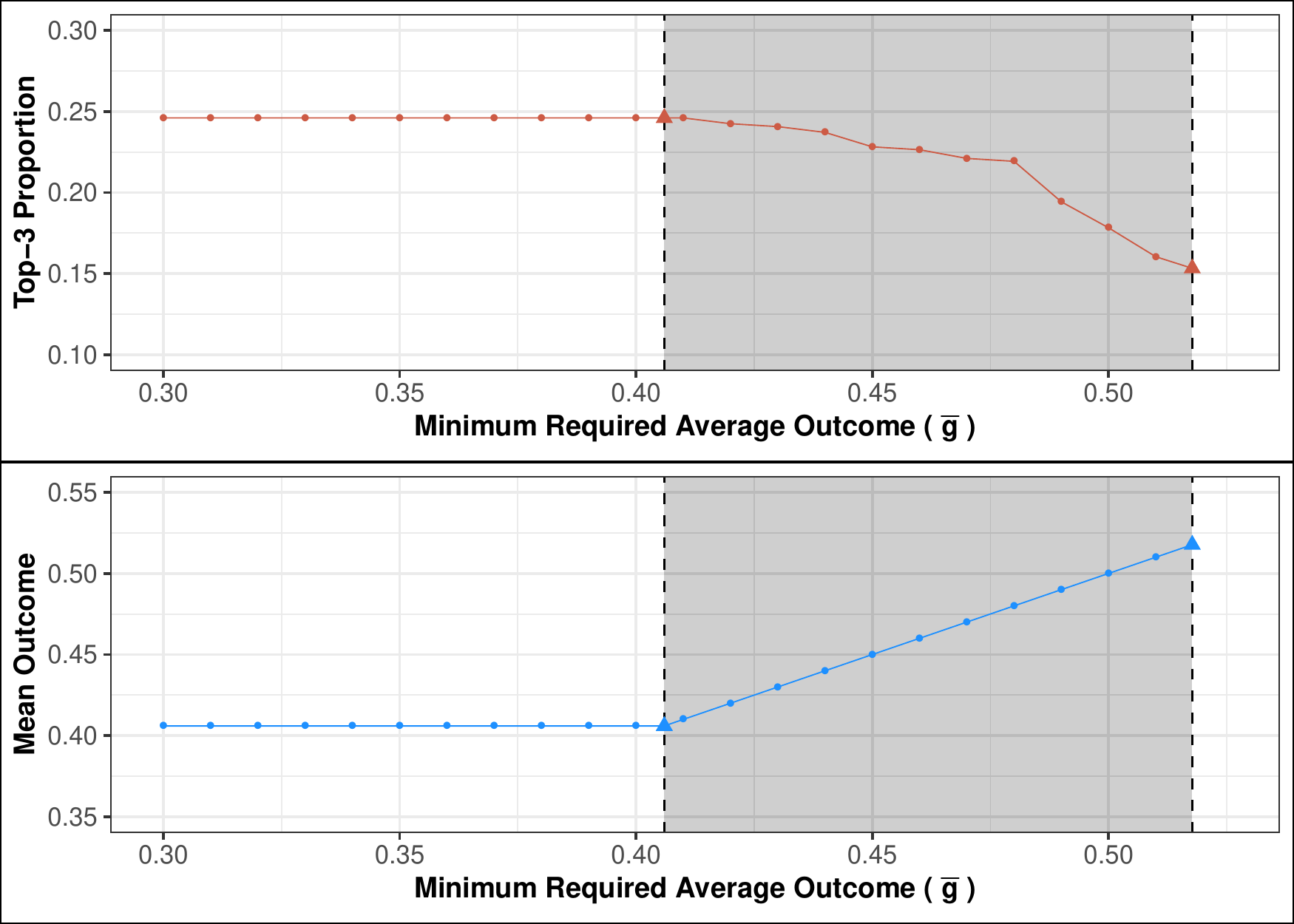}
\caption{\small Results of applying the $\bar g$-Constrained Priority Mechanism to refugee families in the United States for various specified thresholds for the expected minimum level of average integration outcomes ($\bar g$). Upper panel shows the average probability that a refugee family got assigned to one of their top three locations. Lower panel shows the realized average integration outcomes, i.e. the average projected probability of employment. N=561 families who arrived in Q3 of 2016.}
\label{fig:3}
\end{figure}

The bottom panel in Figure \ref{fig:3} shows the distribution of all pairwise correlations between families' outcome vectors.  As can be seen, the correlations are overwhelmingly positive (with a mean correlation of $0.75$), highlighting the fact already shown elsewhere \citep{bansak2018improving} that certain locations are generally better than other locations for helping refugees to achieve positive employment outcomes. However, the fact that there still is meaningful variation across different families' outcome score vectors indicates that certain locations do indeed make a better match for different refugee families, depending on their personal characteristics, which is the foundation for the outcome-optimization matching procedure introduced by \cite{bansak2018improving}.

In applying our mechanism to the 2016 Q3 refugee data, we impose real-world assignment constraints, giving each location capacity for the same number of families as were sent to those locations in actuality. We also truncate each family's preference vectors such that only the first 10 ranks are retained and indifference is established among the remaining locations.

Figure \ref{fig:3} displays the results of applying our mechanism. As before, the mechanism is applied at various levels of $\bar g$, which is denoted by the $x$-axis. The $y$-axis of the top panel denotes the proportion of cases assigned to one of their top three locations, while the $y$-axis in the bottom panel denotes the mean realized outcome score, i.e. the average predicted probability of employment, based on the assignment. The two dashed vertical lines highlight the tradeoff interval, where altering the value of $\bar g$ impacts both preferences and outcomes, and the interval ends when $\bar g$ is raised above $\bar g^{\max}$. 

Given a predominantly preference-based assignment (i.e. setting $\bar g$ to any value below the value at which the tradeoff interval begins), a mean outcome score of $0.41$ is achieved, meaning the predicted average employment rate is 41\%.\footnote{Setting $\bar g$ to a value below the tradeoff interval does not result in a purely preference-based assignment given the trailing indifferences in the preference rank vectors. We also applied the mechanism to the same data without truncating the preference vectors. The result is a purely preference-based assignment at the lowest values of $\bar g$, which yields a mean outcome score of $0.36$. See the SI.}  Under this assignment, about 25\% of refugee families are assigned to a location that is among their top three choices. For comparison, the average observed employment rate for families at their actual locations without applying any mechanism was 34\%. This suggests that there are significant synergies between refugees and locations in the sense that certain locations are a better match for different refugees, depending on their personal characteristics. Even under a predominantly preference-based assignment, the mechanism can therefore increase the predicted average employment rate to $41\%$, about a 20 percent increase over the mean employment rate observed without applying any mechanism. 

On the opposite end of the spectrum, a purely outcome-driven optimization would yield the highest feasible $\bar g$ ($\bar g^{\max}$), which is just below $0.52$, i.e. a predicted average employment rate of 52\%.\footnote{The fact that it is not possible to raise $\bar g$ even further is, of course, the result of the full distribution of the refugee families' outcome vectors, namely the fact that they feature a large positive correlation with one another.} This amounts to about a 53 percent increase over the mean employment rate observed without applying any mechanism. Therefore, if all the government cared about for the assignment was to maximize the score it could attain a considerably higher predicted employment rate by enforcing $\bar g^{\max}$. Yet, at $\bar g^{\max}$, only 15\% of refugee families would be assigned to one of their top three locations. The preference curve in the top panel features a gradient that more gradually steepens, with the tradeoff becoming increasingly more severe as $\bar g$ is increased.

Finally, we also employed an alternative method to estimate location preferences that attempts to correct for potential bias due to relocation costs. As described, we are inferring location preferences from outmigration behavior. However, outmigration decisions are a function of two primary components: a family's desire to leave and their ability to leave. It is the former component that captures preferences and hence what is of primary interest for our purposes, but it is possible that differential costs of leaving and relocating across different locations have an effect on outmigration patterns via the latter component. With only observational behavioral data, is it difficult to perfectly decompose these two components. However, we attempt to do so by estimating a structural model of outmigration designed to capture geographic and economic factors that relate to the costs of relocation, and then by using this structural model to adjust our original preference estimates such that our new preference estimates are, in theory, driven more strictly by the preference component of outmigration behavior. We then re-apply our mechanism to the 2016 Q3 refugee data with the new preferences substituted in. The results, which are provided in the SI, display a similar pattern as when the original preference estimates are employed with one main difference: the proportion of families assigned to one of their estimated top-3 locations is systematically lower at all levels of $\bar g$, which is the result of the families' top-ranked locations being more rivalrous (more highly correlated) according to the alternative preference estimates. More details about the methodology and the results are provided in the SI.

\subsection{Education Data}

We now turn to the results for the application of the mechanism to the education data from Tennessee where we assigned students to elementary schools to optimize on test scores and students' preferences over schools.

\begin{figure}[t!]
\centering
\includegraphics[width=0.85\linewidth]{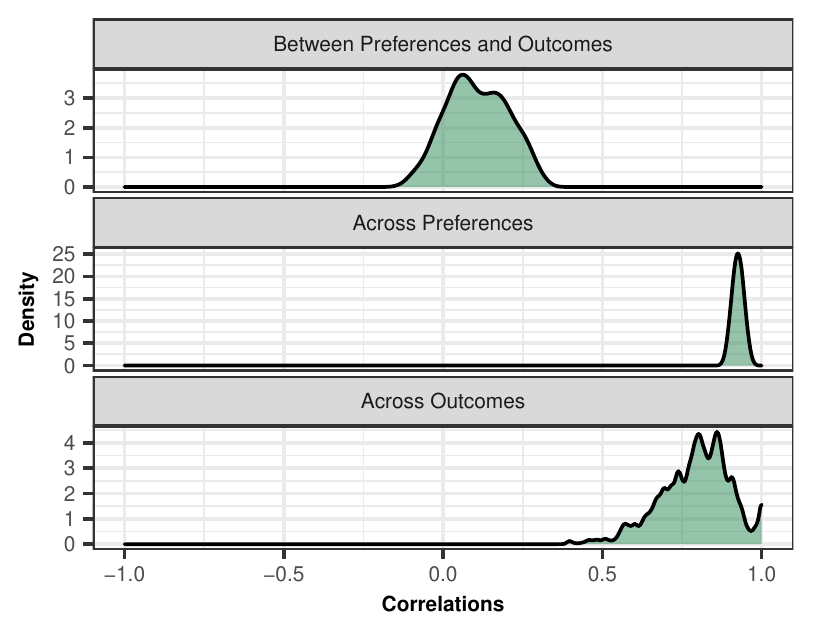}
\caption{\small Shows the distribution of pairwise correlations between student preferences over elementary schools, test score outcomes, and preferences and outcomes. N=1000 randomly sampled students from Tennessee Project Star data.}
\label{fig:Edu1}
\end{figure}

Figure \ref{fig:Edu1} shows features of the joint distribution of the students' outcome score and preference rank vectors. The top panel pertains to the correlation between the students' outcome and preference vectors. We see that for most students the correlations are modest but positive with a mean of $0.11$, indicating that the students slightly prefer schools where they are predicted to have higher test scores. As mentioned earlier, a positive correlation between preference and outcome vectors somewhat alleviates the tradeoff between outcome- and preference-based matching. That said, as shown in the middle panel in Figure \ref{fig:Edu1}, the distribution of pairwise correlations between students' preference vectors are tightly clustered around a high positive mean correlation of $0.93$. This shows that students mostly prefer to be placed in similar schools, which makes the preference-based matching assignment more rivalrous given a fixed number of seats in the preferred schools. As shown in the bottom panel in Figure \ref{fig:Edu1} we also find that the pairwise correlations between students' outcome vectors are almost all positive (with a mean correlation of $0.79$), which suggests that certain schools are generally better than other schools for students to achieve higher test scores.

In applying our mechanism to these education data, we impose the same real-world assignment constraints as before, giving each school capacity for the same number of students as were enrolled in those schools in actuality. We also truncate each student's preference vectors such that only the first 10 ranks are retained and indifference is established among the remaining schools in order to mimic a situation on an application form where students can rank only the top ten preferred schools.

\begin{figure}[t!]
\centering
\includegraphics[width=1\linewidth]{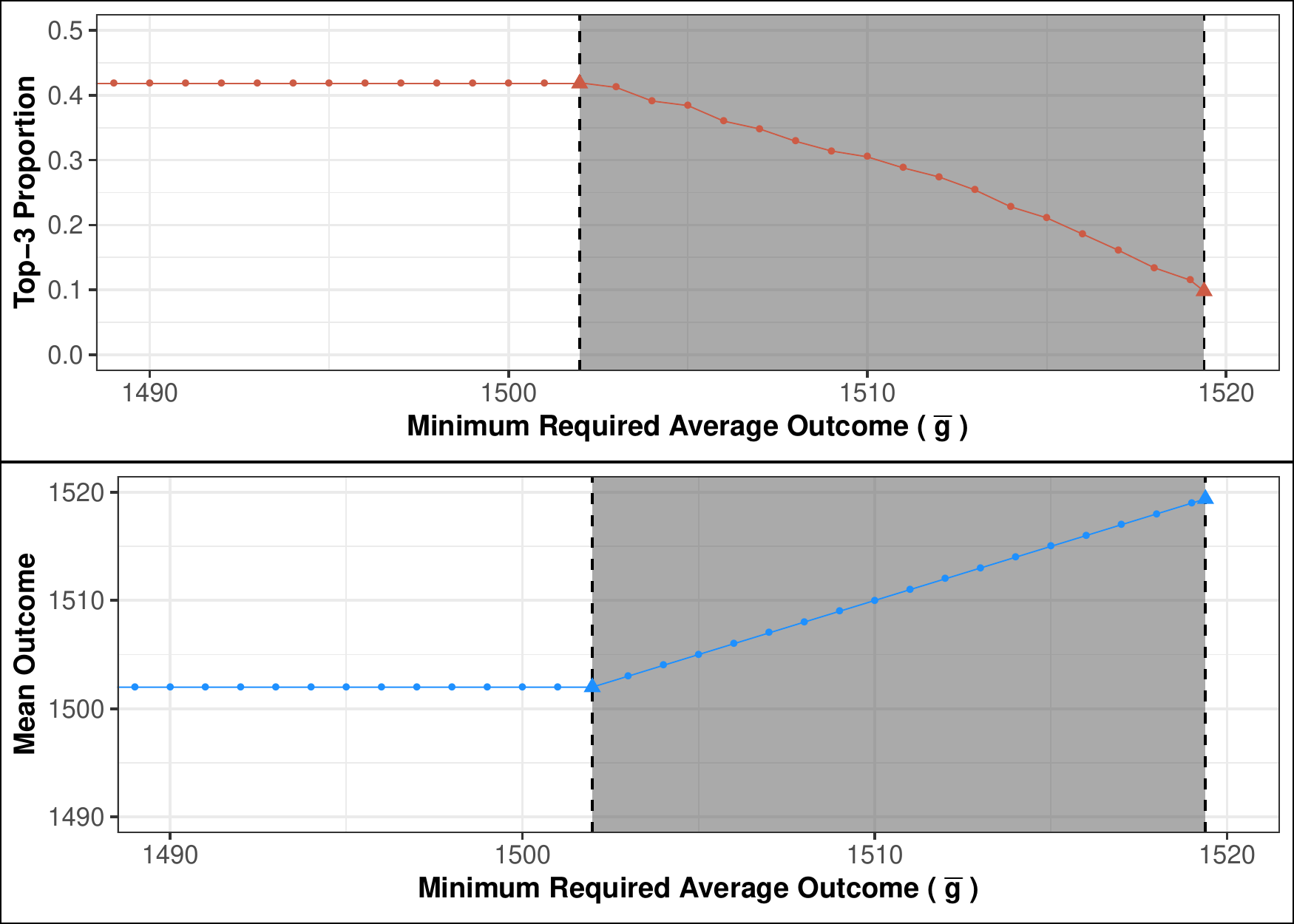}
\caption{\small Results of applying the $\bar g$-Constrained Priority Mechanism to student assignment to elementary schools for various specified thresholds for the expected minimum level of average test score outcomes ($\bar g$). Upper panel shows the average probability that a student got assigned to one of their top three schools. Lower panel shows the realized average test score outcomes, i.e. the average projected SAT score. N=1000 randomly sampled students from Tennessee Project Star data.}
\label{fig:Edu2}
\end{figure}

Figure \ref{fig:Edu2} displays the results of applying our mechanism. As before, the mechanism is applied at various levels of $\bar g$, which is denoted by the $x$-axis. The $y$-axis of the top panel denotes the proportion of students assigned to one of their top three schools, while the $y$-axis in the bottom panel denotes the mean realized outcome score, i.e. the average predicted test score, based on the assignment. The two dashed vertical lines highlight the tradeoff interval, where altering the value of $\bar g$ impacts both preferences and outcomes, and the interval ends when $\bar g$ is raised above $\bar g^{\max}$. 

Given a predominantly preference-based assignment (i.e. setting $\bar g$ to any value below the value at which the tradeoff interval begins), a mean predicted test score outcome of $1502$ is achieved. Under this assignment, about $42\%$ of students are assigned to a school that is among their top three choices.\footnote{Note that this fraction is not directly comparable to the refugee example above since there are a different number of persons, locations, and seats per location.}  For comparison, the average observed test score for the students at their actual locations without applying the mechanism was about $1490$ with a standard deviation of $86$. This suggests that, as with the refugee data above, there are significant synergies between students and schools in the sense that certain schools are a better match for different students, depending on their personal characteristics. Even under a predominantly preference-based assignment, the mechanism can therefore increase the predicted average test score to $1502$, a meaningful improvement of about a seventh of a standard deviation in test scores compared to the observed mean under the actual assignments. 

On the opposite end of the spectrum, a purely outcome-driven optimization would yield the highest feasible $\bar g$ ($\bar g^{\max}$), which is a mean predicted test score outcome of $1519$. A fully outcome-based matching of students to schools can therefore result in a sizable increase in the predicted average test score of about a third of a standard deviation in test scores compared to the observed mean under the actual assignments. Given the tradeoff between preference-based and outcome-based matching, this means that under a purely outcome-driven optimization only about $10\%$ of students would be assigned to a school that is among their top three choices. This highlights that compared to the refugee application, the tradeoff in this education example is somewhat more severe, which is expected given that the preferences are more concentrated on similar schools even though there is a somewhat more positive correlation between preferences and outcomes.

\section{Other Welfare Concerns} 

One possible concern with our mechanism is that if agents whose preferences are not highly correlated with their outcome scores are given higher priority than others, then their assignments could lower the overall preference rank of locations assigned to agents who have lower priority. As an example, besides worrying about achieving a constrained Pareto efficient allocation, suppose the planner also cares about the percentage of agents who are awarded one of their top three ranked locations. Let us refer to this welfare metric as the ``top-3 metric.'' Just how much improvement on this metric can be achieved by changing the order in which families are assigned? 

To get a sense of this, we took a random sampling of the different possible orderings of agents and study the variation generated in the top-3 metric. We re-ran a subset of the nine simulation scenarios considered in Section \ref{sec: simul}, generating the data using identical procedures and employing the same parameters (number of agents, number of locations, size of indifference sets, levels of $\bar g$). However, at each level of $\bar g$ considered in each scenario, we apply the mechanism to the simulated data $100$ separate times where the order of the agents is re-randomized each time. The results are shown in the SI. With respect to the proportion assigned to a top-3 location, the difference between the maximum and minimum ranges from 0.05 to 0.18 with a median difference of 0.13.\footnote{With respect to the mean outcome score, the difference between the maximum and minimum ranges from 0.00 to 0.06 with a median difference of 0.04.} Thus, reordering could produce a typical improvement on the top-3 metric over the typical draw by several percentage points in these data. 

One limitation of this exercise, however, is because the $\bar g$-constrained priority mechanism does not characterize the set of constrained Pareto efficient assignments (as we showed by example in Section \ref{sec: properties}), we do not know if there are constrained efficient assignments that yield improvements even beyond the ones we can generate by re-ordering the families and applying our mechanism. We also do not know if there is a strategy-proof constrained-efficient mechanism that picks out the assignment that maximizes the top-3 metric among those that can be generated by re-ordering the agents under our constrained priority mechanism---let alone an assignment that cannot be generated by re-ordering. It is obvious that the mechanism defined by successively re-ordering and then selecting the one that maximizes the top-3 metric does not define a strategy-proof mechanism. If the planner is willing to sacrifice strategy-proofness, she could attempt to target the best assignment that could be generated using our constrained priority mechanism by successively re-ordering the agents. But by implementing this approach, agents may have an incentivize to falsify their preferences and hence the best assignment(s) the planner is trying to target may no longer even be generated.

\paragraph{Using Predicted Preferences.} An alternative approach to capturing part of the gains that we are seeing from reordering the agents that does not sacrifice either strategy-proofness or constrained efficiency is to use historical (or other) data to predict the preferences of the agents. For example, the planner could use historical data to predict preferences based on the demographic similarities of past agents to current ones, and fix the ordering of agents to be the one that maximizes the top-3 metric according to the predicted preferences.\footnote{As an example, suppose in the refugee matching application that based on historical data, we know that male agents in their 30's that come from a particular country and have worked in a particular profession, are more likely than others to report certain locations as being their top choices. Then we have some noisy prediction of their preferences that in expectation will be correlated with that particular agent's true preferences.} In particular, note that if the planner can perfectly predict the preferences of agents, then strategy-proofness is not a concern since the planner already has the agents' preferences. In this case, the planner can recover the full gain from re-randomizing the order of agents. If the planner cannot perfectly predict the preferences of the agents, but can come close, then the planner should, in expectation, be able to recover some of this gain. Note that because we are using historical data from past agents to set the order of the current agents, the agents cannot use their reported preferences to manipulate the mechanism. 

\begin{figure}[t!]
\centering
\includegraphics[width=1\linewidth]{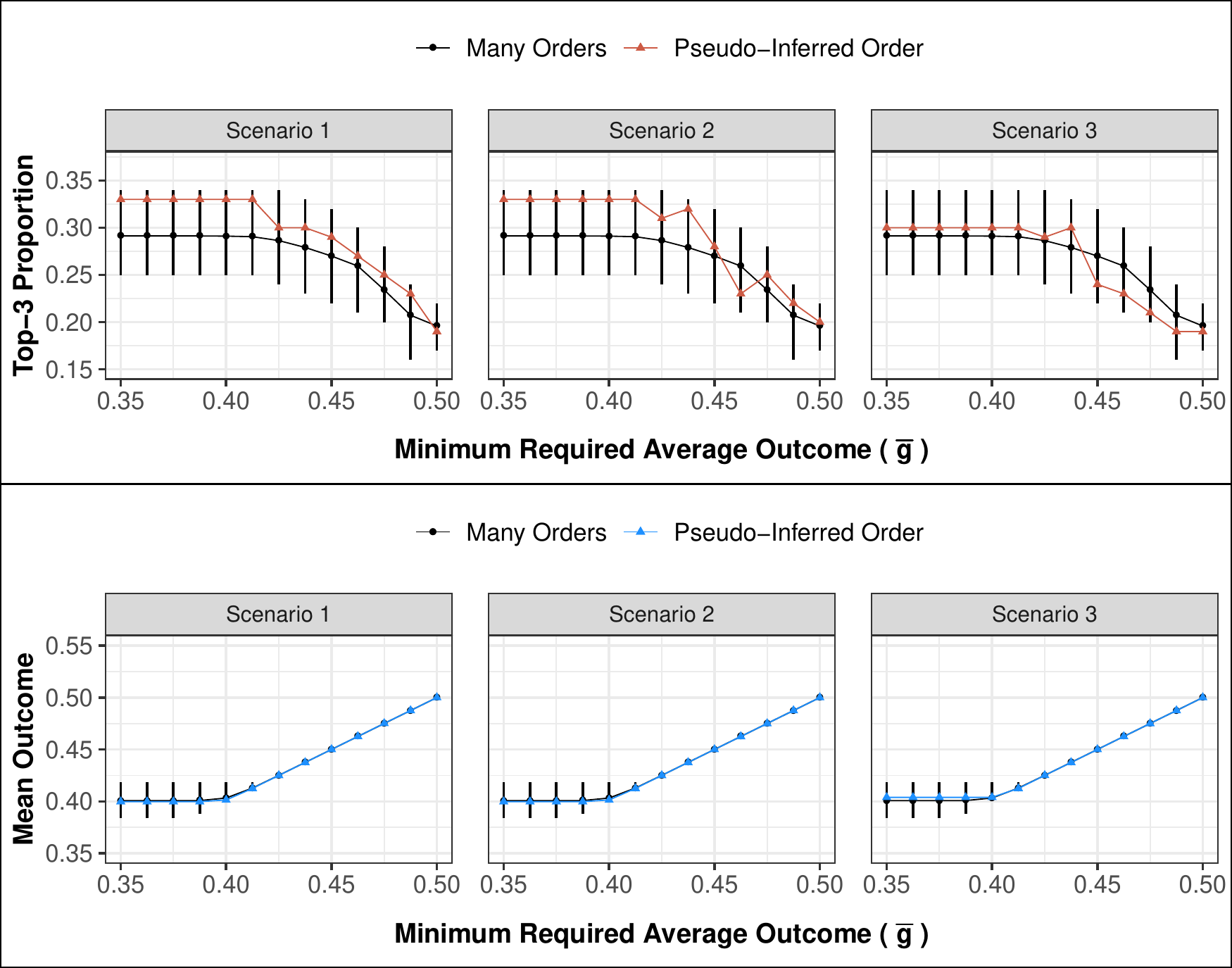}
\caption{\small Results from applying our $\bar g$-Constrained Priority Mechanism to 100 random orderings of a random sample of 100 families from the 2016 Q3 refugee data, along with ``best guess" ordering based on pseudo preferences. The black dots correspond to the average results across the 100 orderings, and the intervals denote the maximum and minimum results obtained across the 100 orderings. The triangles (labeled ``Pseudo-Inferred Order") denote the actual results when employing the ordering that yielded the best pseudo top-3 metric according to the pseudo preferences. The three scenarios successively increase the amount of perturbation applied to the actual preference vectors to generate the pseudo preferences. Upper panel shows the average probability that a agent was assigned to one of its top three locations. Lower panel shows the realized average outcome score. N=100.}
\label{fig:pseudoprefs}
\end{figure}

In order to evaluate this approach, we employed the data from our refugee application described earlier. Specifically, we began by randomly drawing 100 families from the full set of data used in the application. We then randomly generated 100 different orderings of those families, and for each ordering, we implemented the constrained priority mechanism along a sequence of values of $\bar g$. We used the families' outcome score vectors and ``actual" preference vectors (i.e. the preference vectors employed in the application presented earlier). This allows us to asses the extent to which different orderings can result in varying levels of the top-3 metric. The components of Figure \ref{fig:pseudoprefs} labeled ``Many Orders" display these results, with the black dots corresponding to the average across the 100 orderings and the intervals denoting the maximum and minimum results obtained across the 100 orderings.

Furthermore, for each of the 100 orderings, we also evaluated the results of applying the constrained priority mechanism using pseudo-preference vectors in place of the actual preference vectors. These pseudo-preference vectors are intended to represent the imperfectly predicted preferences of the agents. At each level of $\bar g$, we identified the random ordering that resulted in the best pseudo aggregate welfare as measured by the fraction of families receiving one of their top-3 locations according to these pseudo preferences. We were then able to assess the actual welfare results (based on the families' actual preferences) of applying these ``best guess" orderings. In order words, we ran through the process by which a researcher could (i) in advance/independent of actual preference reporting employ simulations to identify orderings likely to lead to higher levels of aggregate welfare based upon predicted preferences, and then (ii) use those as the final orderings by which to actually apply the $\bar g$-constrained priority mechanism to assign the families.

To simulate the process of imperfectly predicting the families' pseudo preferences, we constructed their pseudo preference vectors by randomly perturbing their actual preference vectors. In order to investigate the performance of this approach across different levels of effectiveness in predicting preferences (i.e.,~the extent to which it is possible to construct pseudo preference vectors that are similar to the actual preference vectors), we imposed varying degrees of perturbation and evaluated the results across those different specifications. The results can be seen in the components labeled ``Pseudo-Inferred Order" in Figure \ref{fig:pseudoprefs}, where the separate panels correspond to scenarios with increasing amounts of perturbation.\footnote{The degree of perturbation can be measured in various ways, but one set of intuitive measurements that correspond to our top-3 metric is the proportion of families for whom 3, 2, 1, or 0 of their true top-3 locations are contained in their pseudo top-3 locations. For each of our scenarios, the following reports the computed proportion of families for whom 3, 2, 1, or 0 of their true top-3 locations are contained in their pseudo top-3 locations. Scenario 1: 0.77 (3), 0.23 (2), 0.00 (1), 0.00 (0). Scenario 2: 0.37 (3), 0.59 (2), 0.04 (1), 0.00 (0). Scenario 3: 0.03 (3), 0.33 (2), 0.51 (1), 0.13 (0).} For each scenario, the triangles labeled ``Pseudo-Inferred Order" denote the actual results when applying the ordering deemed best according to the pseudo preference results, as described above. The figure shows that a large portion of the gain from carefully fixing the order of agents could be recovered if the planner is able to very accurately predict preferences, but how much can be gained could be very sensitive how good a prediction the planner makes.\footnote{In the graphs on the far right, the planner is able to predict at least one of the top three locations for more than half the agents, and at least two for a third of them.} 

\paragraph{Ordering Agents by Outcome Score Variance.} We also explored an alternative strategy for identifying \emph{a priori} (and hence without sacrificing strategy-proofness) an agent ordering that is likely to result in a favorable level of the top-3 metric. Rather than attempting to predict pseudo preferences, this strategy instead utilizes the families' outcome scores. Specifically, for each family the variance of outcome scores across locations can be computed, and then agents can be ordered according to their variances. We propose ordering the agents in increasing variance (from lowest variance to highest variance). The intuition for this proposal is the following. In making assignments, the $\bar g$-constrained priority mechanism is faced with the tradeoff between sending agents to their preferred location and sending them to a location that will enable the $\bar g$ constraint to be met. For each agent, the extent to which this particular tradeoff can bite depends to some degree on the variance of their outcome scores across locations. In the extreme case, there is no tradeoff for an agent whose outcome score is identical across locations: no matter where they are sent, their assignment will have an equivalent implication for the $\bar g$ constraint. However, for agents whose outcome score variance is very high, the extent to which their assignment can work in favor of or against the $\bar g$ constraint varies greatly across locations. In other words, the high-variance agents' assignments offer opportunity to create buffer for the $\bar g$ constraint, whereas the low-variance agents' assignments do not. Therefore, assigning low-variance agents earlier ensures that such units are more likely to be assigned to a highly preferred location without occurring at the expense of excessively cutting into the $\bar g$ constraint.\footnote{Another way to view this is that because the assignment decision for the high-variance agents has more influence on the $\bar g$ score---and hence their assignment also offers more potential to create buffer against violating the $\bar g$ constraint---then from the perspective of managing the tradeoff between preferences and outcome scores, it does not make sense to waste the potential these units offer by assigning them early, given that earlier assignments will more strongly prioritize preferences.}

\begin{figure}[t!]
\centering
\includegraphics[width=0.75\linewidth]{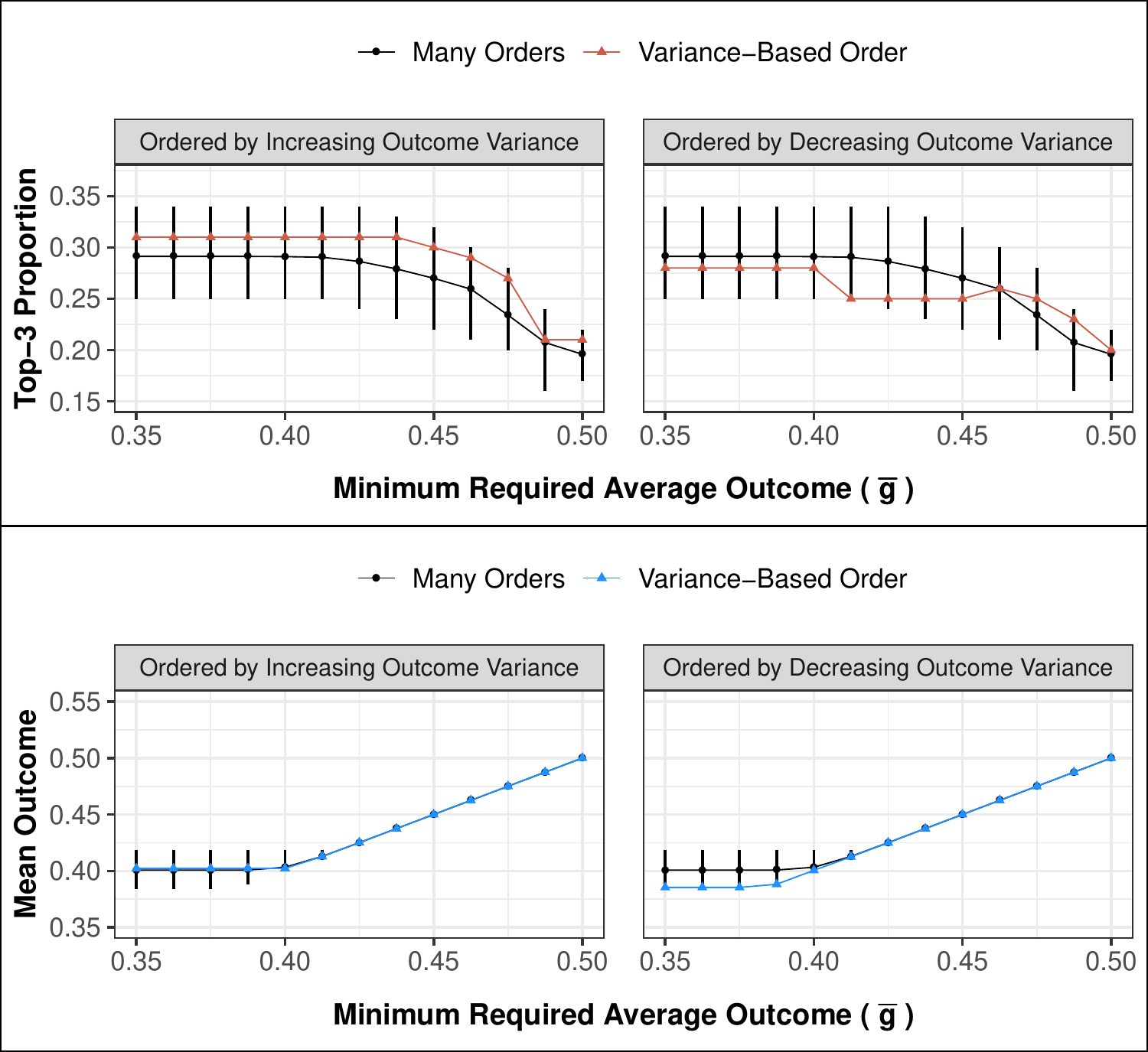}
\caption{\small Results from applying our $\bar g$-Constrained Priority Mechanism to 100 random orderings of a random sample of 100 families from the 2016 Q3 refugee data, along with outcome score variance-based orderings. The black dots correspond to the average results across the 100 orderings, and the intervals denote the maximum and minimum results obtained across the 100 orderings. The triangles (labeled ``Variance-Based Order") denote the results when employing orderings based on the families' outcome score variances across locations, with families ordered by increasing variance on the left and decreasing variance on the right. Upper panel shows the average probability that a agent was assigned to one of its top three locations. Lower panel shows the realized average outcome score. N=100.}
\label{fig:variancebasedorders}
\end{figure}

Figure \ref{fig:variancebasedorders} shows the results of applying this increasing-variance ordering strategy (left panels), as well as the opposite decreasing-variance ordering for illustrative purposes (right panels). The components labeled ``Many Orders'' display the results from the same 100 random orderings as in Figure \ref{fig:pseudoprefs}, with the black dots corresponding to the average across the 100 orderings and the intervals denoting the maximum and minimum results obtained across the 100 orderings. The components labeled ``Variance-Based Order" display the results when applying the mechanism to the families put in the proposed increasing-variance order (left panels), or in the decreasing-variance order (right panels). The figures show that when agents are ordered by increasing outcome score variance, a substantial  share of the gain from fixing the order of agents can be recovered.  It also depicts what we expect would happen when agents are ordered by decreasing outcome score variance, that welfare measured by the top-3 metric is generally worse than under the typical random ordering. 

\paragraph{Overview.} One property of the increasing outcome variance ordering approach is that it does not rely on historical (or any other) data to predict the preferences of agents. Another desirable property of this approach relates to fairness/distributive considerations: the agents that cannot gain much by way of their outcome score through different assignments can be prioritized to achieve gains in terms of their preferences, and those agents that have a lot to gain by way of outcome scores could enjoy such gains even if they do not get one of their top preferences. A concern with fixing the order of agents using either of these techniques, however, is that it may unintentionally though systematically put agents of particular backgrounds higher or lower in the priority order, a form of disparate impact that may not be desirable to the planner. In choosing the order of agents, one might want to incorporate additional constraints to overcome part of the bias that may be implicit in these techniques. More theoretical work will be necessary to evaluate how workable these and other approaches of using data and machine learning techniques to determine the order of agents will be in various applications.

\section{Other Mechanisms}

As we have shown, the priority mechanism is a mechanism for which we can add a welfare constraint without compromising important properties such as strategy-proofness and (constrained) efficiency. One could ask whether we can amend other existing mechanisms to take into account the same constraint while retaining their desirable properties.

\paragraph{Top Trading Cycles.} 

One candidate for an alternative mechanism, among matching mechanisms with one-sided preferences, is Gale's Top Trading Cycles (TTC) mechanism \citep{shapley1974cores,roth1982incentive}, which has already been employed in previous proposals for refugee matching \citep{delacretaz2016refugee}. However, adding a planner's $\bar g$ constraint to this mechanism while retaining the feature that it is strategy-proof and constrained efficient is not straightforward. 

\begin{figure}[th!]
\centering
\includegraphics[scale=.3]{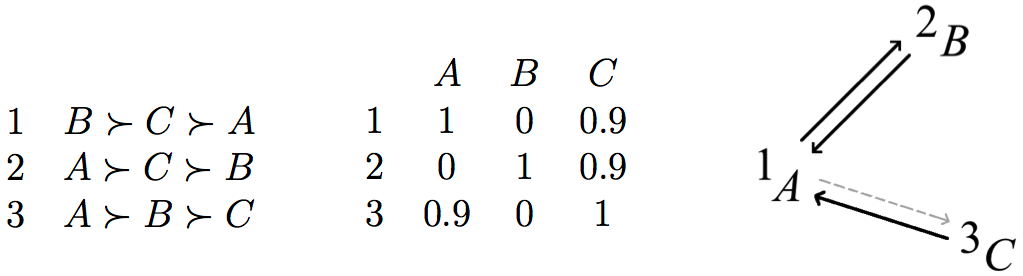}
\caption{\small Three person-three location example showing violation of strategy-proofness when adding a planner's $\bar g$ constraint to TTC mechanism.}
\label{fig:fig_ttc}
\end{figure}

Consider, for example, the simple adjustment of this mechanism that begins by provisionally assigning agents to locations to maximize the planner's objective, and removes cycles until the planner's welfare measure falls below the threshold, at which point it stops and everyone that is unassigned receives the assignment that they currently provisionally have. The following three person-three location example depicted in Figure \ref{fig:fig_ttc} shows that this mechanism is not strategy-proof.  The agents are $1, 2, 3$ and locations are $A$, $B$, $C$. 

To maximize the average outcomes score, agent 1 is provisionally assigned to $A$, 2 to $B$, and 3 to $C$. Preferences over locations are given on the left. In the middle, we have the values of $g_i(l)$. Under truthful reporting agent 1 points to $B$, and 2 and 3 point to $A$. The only cycle is between $1$ and $2$. However, if the planner's threshold $\bar g$ is set to 0.5, then swapping 1 and 2's locations guarantees an average outcome score below this threshold. The algorithm would terminate with the outcome maximizing assignment being assigned. However, agent 1 could do better by misreporting and pointing to $C$ instead. In this case, the assignment would be 1 to $C$, 2 to $B$, and 3 to $A$, which gives an average outcome score above the threshold.

Thus, while it may be possible to incorporate an outcome constraint into the TTC mechanism that preserves strategy-proofness and constrained efficiency, it appears that there is no straightforward way to do so. For the priority mechanism, however, incorporating this constraint is both straightforward and computationally tractable.

\paragraph{Two-Sided Mechanisms.}

Finally, we could also consider matching mechanisms with two-sided preferences such as the deferred acceptance mechanism \citep{gale1962college} where we incorporate the planner's welfare objective into the preferences for the locations. Here, there are at least two possibilities. First, we could assume that locations care about maximizing the planner's welfare score, along with other considerations; that is, we allow the locations to express their genuine preferences. At least for the refugee assignment application, this appears to be politically challenging, partly because policymakers are concerned that this could result in political problems where some locations might discriminate against refugees from certain groups/nationalities.  The second possibility is we assume that each location simply wants to maximize the average outcome score among agents assigned to it. This creates competition among locations. Again, at least in the refugee assignment application, it is not clear why the planner (national government) would want to allow this---i.e. it is not clear what this assignment mechanism would achieve that serial priority does not, given the objectives of the planner. 

\section{Conclusion}

We have proposed an assignment mechanism for contexts where there is a social planner/designer with their own welfare objective. Our mechanism strikes a compromise between maximizing the planner's objective and conducting the assignment solely on the basis of the agents' preferences. The mechanism is strategy-proof, constrained efficient, and does not require agents to rank all locations. In real-world implementations of our mechanism, a planner could either fix a feasible value of $\bar g$ in advance or review the projected results along a sequence of $\bar g$ values (as in Figures \ref{fig:3} and \ref{fig:Edu2}) and choose the final preferred assignment according to their own criteria.

We applied our mechanism to refugee assignment and school choice data to demonstrate how it could be implemented. Refugee matching has become a prominent policy innovation proposed to help facilitate the successful integration of refugees into host countries' economies and societies. However, there is disagreement over whether integration is best served by matching on refugee preferences or expected integration outcomes. Our study highlights the value for governments to collect preference information from refugees to provide them with agency and improve allocations by harnessing the value of private information they possess over which locations work best for them. In addition, our mechanism is applicable to other domains that involve the assignment of agents to different types of locations (or more generally speaking, one-to-one and many-to-one bipartite matching problems). As a second example, we apply our mechanism to the assignment of kindergarteners to schools. School choice has been a longstanding application of market design, and our illustration demonstrates how our mechanism can be applied to this canonical setting.

In addition, our investigation resulted in interesting new theoretical insights. First, we discovered that the priority mechanism appears to be unique in the sense that our outcome constraint can be incorporated into it in a straightforward manner without sacrificing the important properties of strategy-proofness, efficiency, and computational tractability. In contrast, the simple modifications of the top trading cycle that we considered to incorporate an outcome constraint did not retain strategy-proofness and/or computational tractability. Future research might consider other modifications that retain these properties. Second, we also discovered that not all of the canonical properties of the priority mechanism are inherited by our constrained version. Namely, the $\bar g$-constrained priority mechanism does not characterize the full set of constrained efficient assignments. 
 
These applications of our mechanism provide examples of how predictive analytics from machine learning can be fruitfully combined with the preference-based allocation schemes common in market design. The marriage of these two approaches can provide a powerful tool to improve allocations in a way that incorporates information about what people want while harnessing the statistical learnings from the historical data about what would be the best options. Given the heterogeneity in information levels and the richness of historical data on outcomes, we envision that such a combined approach could lead to better allocations in a variety of settings compared to schemes that rely only on preferences or only on expected outcomes.

\section*{Acknowledgments}

We acknowledge funding from the Rockefeller Foundation, Schmidt Futures, and the 2018 HAI seed grant program from the Stanford AI Lab, Stanford School of Medicine, and Stanford Graduate School of Business. The funders had no role in the data collection, analysis, decision to publish, or preparation of the manuscript. We thank the Lutheran Immigration and Refugee Service for access to data and guidance. We are grateful to Fuhito Kojima and Shunya Noda for help and guidance.

\section*{Data Availability Statement}

Replication code for this article has been published in Code Ocean, a computational reproducibility platform that enables users to run the code, and can be viewed interactively at \cite{codeocean2020abh} or \url{https://doi.org/10.24433/CO.3735899.v1}. A preservation copy of the same code and data can also be accessed via Harvard Dataverse at \cite{DVN/ZEV0WX} or \url{https://doi.org/10.7910/DVN/ZEV0WX}. The U.S. refugee data were provided to us under a collaboration research agreement with the Lutheran Immigration and Refugee Service (LIRS). This agreement requires that we do not transfer or disclose the data. Researchers interested in the data can contact LIRS at 700 Light Street, Baltimore, Maryland 21230, \textsf{lirs@lirs.org}. We declare that we have no competing interests.

\bibliography{references}

\clearpage

\appendix

\begin{center}
\huge \textbf{Supplemental Information (SI)} \\ \vspace{3.75cm} \LARGE Combining Outcome-Based and Preference-Based Matching: A Constrained Priority Mechanism \\ \large \vspace{1.5cm} Avidit Acharya, Kirk Bansak, Jens Hainmueller
\end{center}

\normalsize

\label{sec: appendix}

\pagenumbering{roman}
\thispagestyle{empty}
\setcounter{page}{0}

\numberwithin{lemma}{section} \setcounter{lemma}{0}
\numberwithin{definition}{section} \setcounter{definition}{0}
\numberwithin{proposition}{section} \setcounter{proposition}{0}
\numberwithin{theorem}{section} \setcounter{theorem}{0}
\numberwithin{lemma}{section} \setcounter{lemma}{0}
\numberwithin{corollary}{section} \setcounter{lemma}{0}
\renewcommand{\thefigure}{S\arabic{figure}}
\renewcommand{\thetable}{S\arabic{table}}
\setcounter{table}{0}
\setcounter{figure}{0}

\clearpage

\section{Properties of the Mechanism}

\subsection{Proof that it is Constrained Efficient}

Suppose that $\varphi$ is not $\bar g$-constrained efficient, so that for some preference profile $\succsim$, $\varphi(\succsim)$ is Pareto-dominated by a feasible $\bar g$-acceptable matching $\mu$. 

For all families $i$, let $M_i = \{j < i: j \notin N_i\}$ be the families ahead of $i$ that were already assigned a location under $\varphi(\succsim)$, and let $\underline i = \min \{i : \mu (i) \succ_i \varphi(\succsim)(i)\}$ be the first family to which $\mu$ assigns it a location that it strictly prefers to the one it gets under $\varphi(\succsim)$. (Such a family must exist if $\mu$ Pareto-dominates $\varphi(\succsim)$.) By construction $\mu(i) = \varphi (\succsim)(i)$ for all $i \in M_{\underline i}$. So for $\mu$ to be feasible and $\bar g$-acceptable, it must be that $\mu(\underline i) \in S_{\underline i} \cap L^{\bar g}_{\underline i} (\alpha_{\underline i})$, where $\alpha_{\underline i}$ is the completed assignment under $\varphi(\succsim)$ at Step $\underline i$. This means that $S_{\underline i} \cap L^{\bar g}_{\underline i} (\alpha_{\underline i}) \neq \emptyset$ so $\varphi(\succsim)$ must have assigned the best location $l^*_{\underline i}$ in this set to family $\underline i$. But since $\mu(\underline i) \succ_i \varphi(\succsim)(\underline i) = l_{\underline i}^*$, this contradicts the assumption that $l_{\underline i}^*$ is the best location for $\underline i$ in $S_{\underline i} \cap L^{\bar g}_{\underline i} (\alpha_{\underline i})$. 

\subsection{Proof that it is Strategy-proof}

Suppose that there is some $i$ for whom reporting a different preference $\succsim'_i$ produces a strictly better location assignment: $\varphi (\succsim'_i, \succsim_{-i}) \succ_i \varphi (\succsim)(i)$. 

Let $l'_i = \varphi (\succsim'_i, \succsim_{-i})$ and note that $S_j \cap L_j^{\bar g}(\alpha_j)$ is independent of $i$'s reported preference for all $j < i$. Therefore, $N_i = N'_i$ where $N_i$ is the set of families on hold at Step $i$ under the truthfully reported profile $\succsim$ and $N'_i$ are those on hold at Step $i$ under the profile $(\succsim'_i, \succsim_{-i})$. In addition, $\varphi (\succsim'_i, \succsim_{-i})(j) = \varphi (\succsim)(j)$ for all $j \in N_i$. This implies that $\alpha'_i = \alpha_i$, where $\alpha'_i$ is the completed assignment at Step $i$ under preference profile $(\succsim'_i, \succsim_{-i})$ and $\alpha_i$ is the completed assignment at Step $i$ under preference profile $\succsim$. Therefore, $L_i^{\bar g}(\alpha_i) = L_i^{\bar g}(\alpha'_i) =: L_i^{\bar g}$. 

Let $S'_i$ be the locations that $i$ ranks strictly under $\succsim'_i$ and $S_i$ the locations that $i$ ranks strictly under $\succsim_i$. If $S_i \cap L_i^{\bar g} = \emptyset$, then all of the locations in $L_i^{\bar g}$ are ones that $i$ ranks worst, and $i$ is guaranteed to be assigned one of these locations regardless of which location $i$ reports. Therefore it cannot be that $\varphi (\succsim'_i, \succsim_{-i}) \succ_i \varphi (\succsim)(i)$. 

On the other hand, if $S_i \cap L_i^{\bar g} \neq \emptyset$ then $\varphi (\succsim'_i, \succsim_{-i}) \succ_i \varphi (\succsim)(i)$ and $L_i^{\bar g}(\alpha_i) = L_i^{\bar g}(\alpha'_i)$ implies that $l'_i \in S_i \cap L_i (\alpha_i)$. But then $l'_i \succ_i \varphi(\succsim)(i) = l_i^*$ contradicts the fact that $l_i^*$ is the unique best location in $S_i \cap L_i (\alpha_i)$ under preference $\succsim_i$.

\section{Verifying $\bar g$-Acceptability}

As described in the main text, implementing the $\bar g$-constrained priority mechanism involves iteratively verifying that the next assignment of a family to a particular location can be performed without compromising the possibility of a $\bar g$-acceptable final matching. This process requires solving the maximization problem in Equation 2 of the main text:
\begin{align}
G_i (q^i) := & \max_{\beta_i} \sum_{j \in \{i,...,n\} \cup N_i} g_j (\beta_i (j)) \text{ subject to } |\beta_i^{-1} (l) | \leq q^i_l,  \forall l  \tag{2}
\end{align}
\noindent
This involves computing the maximum possible total outcome score for any remaining set of units and the remaining location capacities. 

In implementing the mechanism, Equation 2 can be solved by employing a standard linear sum assignment problem (LSAP) \citep{burkard2012assignmentch4}. Specifically, the LSAP formulation is applied to an augmented cost matrix, whereby the rows correspond to the remaining units and the columns correspond to location capacity slots (i.e. each column is replicated according to the number of capacity slots belonging to the associated location). Each element $[i,v]$ of the cost matrix corresponds to the complement of the outcome score for the $i$th unit when assigned to the location to which the $v$th column pertains.

Various polynomial-time algorithms have been developed for solving the LSAP, beginning with the introduction of the Hungarian algorithm in the 1950s \citep{kuhn1955hungarian, munkres1957algorithms}. We employ the RELAX-IV cost flow solver developed by Bertsekas and Tseng \citep{bertsekas1994relax} and implemented in \texttt{R} by the \texttt{optmatch} package \citep{hansen2006optimal}.

\section{Simulation Application: Additional Details}

The follow describes the data-generating process employed in the simulations. 

First a number $N$ is chosen, denoting the number of agents. For simplicity, the same number of locations is also used, each with capacity for one agent. In addition, $\rho_p$ and $\rho_{op}$ are both chosen, denoting the pre-specified correlation between preferences across agents and the correlation between preferences and outcome scores within agents.

Next, $N$ different $N$-dimensional latent variable vectors are generated, and these vectors are column-bound into an $N$ x $N$ matrix, which we denote by $\textbf{P}$, representing a simulated preference matrix. Specifically, each vector is a multivariate normal random vector, using a mean vector of $0$, and a covariance matrix with $1$ for all the diagonal elements and $\rho_p$ for all the off-diagonal elements. Let $\vec{z}_l$ denote the $l$th $N$-dimensional latent variable vector, which pertains to the $l$th location and comprises the $l$th column of $\textbf{P}$. For any given vector, the $i$th element pertains to the $i$th family.

By generating the $N$ x $N$ matrix $\textbf{P}$ in this way, each row represents a client and each column represents a location. Thus, the $i$th row, $\textbf{P}$[i,], denotes a latent preference vector for agent $i$, with higher (more positive) values corresponding to a higher preference and vice versa. By construction, for any two cleints (rows), the pairwise correlation between the two vectors will be $\rho_p$ in expectation, imposing a correlation of $\rho_p$ across agents' preferences over locations.

Let $\vec{s}_i$ denote the $i$th cleint's outcome score vector. The outcome score vectors are constructed such that $\vec{s}_i = sign( \rho_{op} ) \cdot (\textbf{P}[i,] + \vec{\epsilon})$, where the elements of $\vec{\epsilon}$ are independently distributed normal with mean 0 and variance $\sigma^2_{\epsilon}$. The value of $\sigma^2_{\epsilon}$ is determined such that it, in combination with the $sign( \rho_{op} )$ operator, produces an expected pairwise correlation of $\rho_{op}$ between $\vec{s}_i$ and $P[i,]$, thereby inducing the correlation of $\rho_{op}$ between a agent's preferences and outcome scores. The outcome score vectors are then row-bound to create an $N$ x $N$ outcome score matrix $\textbf{S}$, where each row represents a agent and each column represents a location. 

In applying our mechanism to the simulated data, the $\textbf{S}$ matrix is first normalized such that its elements are in the interval $[0,1]$, and the $\textbf{P}$ matrix is mapped to preference ranks (i.e. each row $\textbf{P}[i,]$ is transformed into ranks such that the most positive value becomes $1$ and the most negative value becomes $N$).

For simplicity, the simulations presented in the study employ $N = 100$ (i.e. $100$ agents assigned to $100$ locations each with one seat). In addition, to mimic reality, in which agents are likely to be able to report only a limited number of location preferences, the preference vectors for each agent are truncated such that only the top $10$ ranks are retained and indifference is established among the remaining locations. The simulations vary both the correlation between preference and outcome vectors (three values of $\rho_{op}$: -0.5, 0, and 0.5) and the correlation between preference vectors across agents (three values of $\rho_{p}$: 0, 0.5, and 0.8). This yields nine different scenarios, and in each we apply our mechanism to make the assignment for various values of $\bar g$. Figure 1 in the main text displays the results.

\begin{figure}[b!]
\centering
\includegraphics[width=1\linewidth]{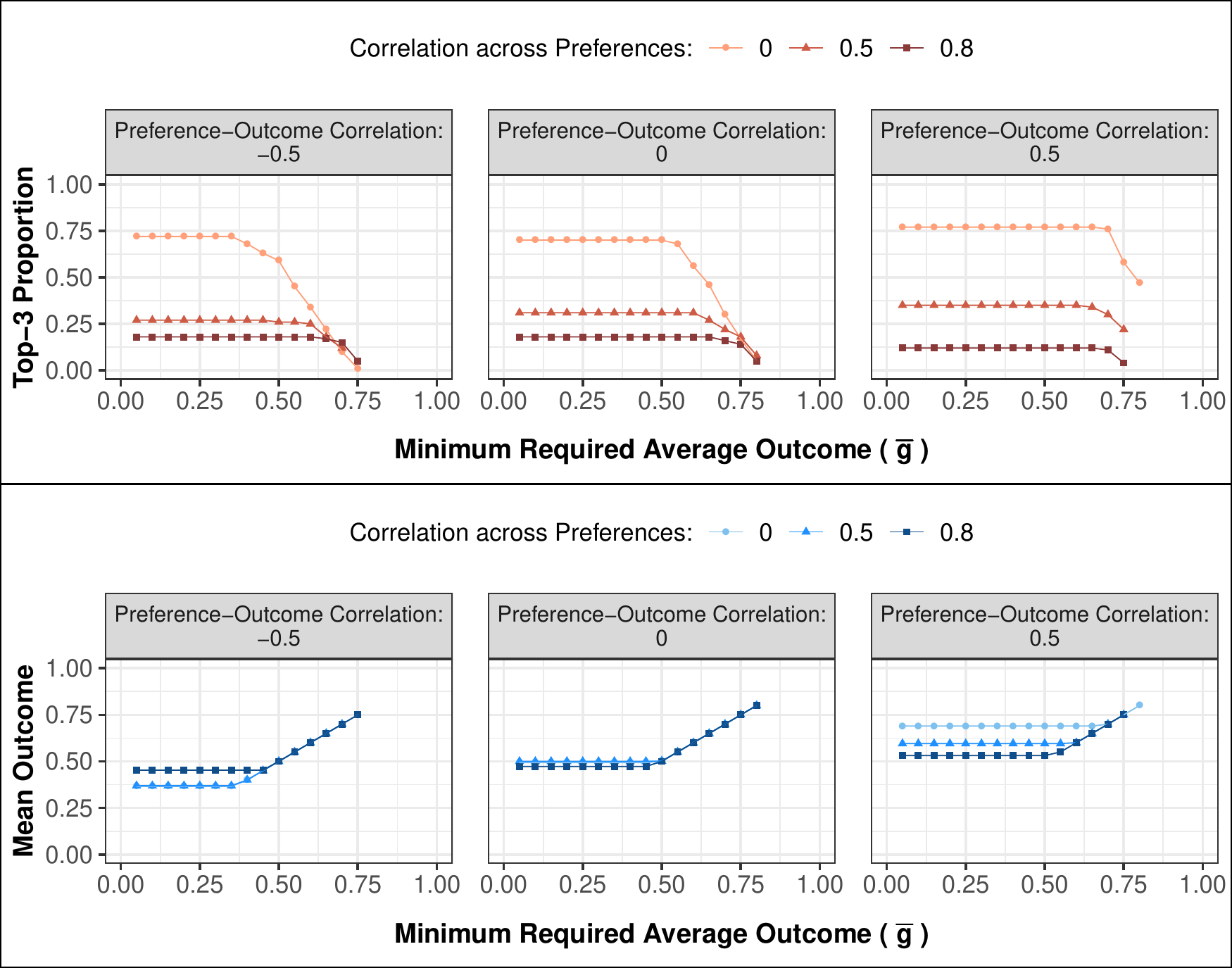}
\caption{Results from applying our $\bar g$-Constrained Priority Mechanism to simulated data (without truncated preferences) that varies the correlations between location preference and integration outcome vectors and the correlations between preference vectors across families. This figure shows the results of the same simulations as in the main text Figure 1, except that the simulated families' preference rank vectors were not truncated in the simulations illustrated here. Upper panel shows the average probability that a family was assigned to one of its top three locations. Lower panel shows the realized average integration outcomes, i.e. the average projected probability of employment. $N=100$.} \label{fig:sim_notruncCPM}
\end{figure}

In addition, Figure \ref{fig:sim_notruncCPM} in this SI shows the results of the same simulations when the preference rank vectors are not truncated.

\section{U.S. Refugee Application}

\subsection{Background Information on U.S. Resettlement}

Resettled refugees in the United States are assigned to locations based on collaboration between the Department of State and nine voluntary resettlement agencies. During a regular draft, refugees are first allocated to one of the nine agencies according to specific quotas. Agencies are then responsible for assigning refugees to locations within their networks. Typically refugees are assigned as cases, where a case is a family. The assignment varies based on whether the refugee has family ties in the United States. Refugees with ties are placed at the location most proximate to the tie. Refugees without such ties, so-called ``free cases,'' are assigned on a case-by-case basis and can be assigned to any location in the network. Placement officers consider special characteristics of the case (nationality, case structure, medical needs) and consult with the local offices on whether they can accommodate a case (e.g. some offices may lack interpreters for particular languages). Among the offices that can accommodate a case, the case is then typically assigned to offices with the smallest proportion of their yearly capacity currently filled. Note that a different process applies to refugees with Special Immigrant Visas (SIVs). 

Once a refugee case has been assigned, the local office then provides placement and reception services for 90 days beginning after arrival as mandated by the U.S. Resettlement Program. The duration is 180 days for refugees assigned to the matching grant program. Agencies are mandated to report employment outcomes to the Department of State after the conclusion of the placement and reception period. If a refugee leaves the area before the placement and reception period ends, they may no longer receive the benefits associated with the placement and reception service. 

\subsection{Registry Data}

Our data includes all refugees that were resettled by one of the largest resettlement agencies and arrived between quarter 1, 2011 and quarter 3, 2016. The same data is used in \cite{bansak2018improving}. We restrict the sample to those aged between 18 and 64 years at the time of arrival (i.e. working age). We also remove a small number of duplicates and locations that have had less than 200 refugees assigned to them over the entire period. In the final data there are 33,782 refugees from 22,144 cases. Of those, 9,506 refugees are from free cases. 

Table \ref{tab:destatUS} shows the descriptive statistics for our sample. Below is a list of variables and measures used:
\begin{itemize}
\item \emph{Male}: Binary variable coded as 1 for males and 0 for females.
\item \emph{Speaks English}: Binary variable coded as 1 for refugees who speak English at the time of arrival and 0 otherwise.
\item \emph{Age at arrival}: Age at arrival measured in years.
\item \emph{Education}: Highest level of educational attainment at arrival. Categories include: None/Unknown, Less than Secondary, Secondary, Advanced, and University.
\item \emph{Country of origin}: Country of origin or nationality.
\item \emph{Employed}: Binary variable coded as 1 for refugees who are employed at 90 days after arrival, and 0 otherwise.
\item \emph{Year of arrival}: Year of arrival (continuous).
\item \emph{Month of arrival}: Month of arrival (continuous).
\item \emph{Free case}: Binary variable coded as 1 for refugees who are free cases with no U.S. ties, and 0 otherwise.
\end{itemize}

\subsection{Applying the Mechanism}

We applied our mechanism to the data on the refugee families who arrived in the third quarter (Q3) of 2016, specifically focusing on refugees who were free to be assigned to different resettlement locations (561 families, 919 working-age individuals). To generate each family's outcome score vector across each of the locations, we employed the same methodology in \cite{bansak2018improving}, using the data for the refugees who arrived from 2011 up to (but not including) 2016 Q3 to train gradient boosted tree models that predict the expected employment success of a family (i.e. the mean probability of finding employment among working-age members of the family) at any of the locations, as a function of their background characteristics. These models were then applied to the families who arrived in 2016 Q3 to generate their predicted employment success at each location, which comprise their outcome score vectors. 

To generate preference rank vectors, we infer revealed location preferences from secondary migration behavior. Specifically, we use the same modeling procedures used in the outcome score estimation, simply swapping in outmigration in place of employment as the response variable. This allows us to predict for each refugee family that arrived in 2016 Q3 the probability of outmigration at each location as a function of their background characteristics. For each family, we then rank locations such that the location with the lowest (highest) probability of outmigration is ranked first (last).

\begin{figure}[tbhp]
\centering
\includegraphics[width=1\linewidth]{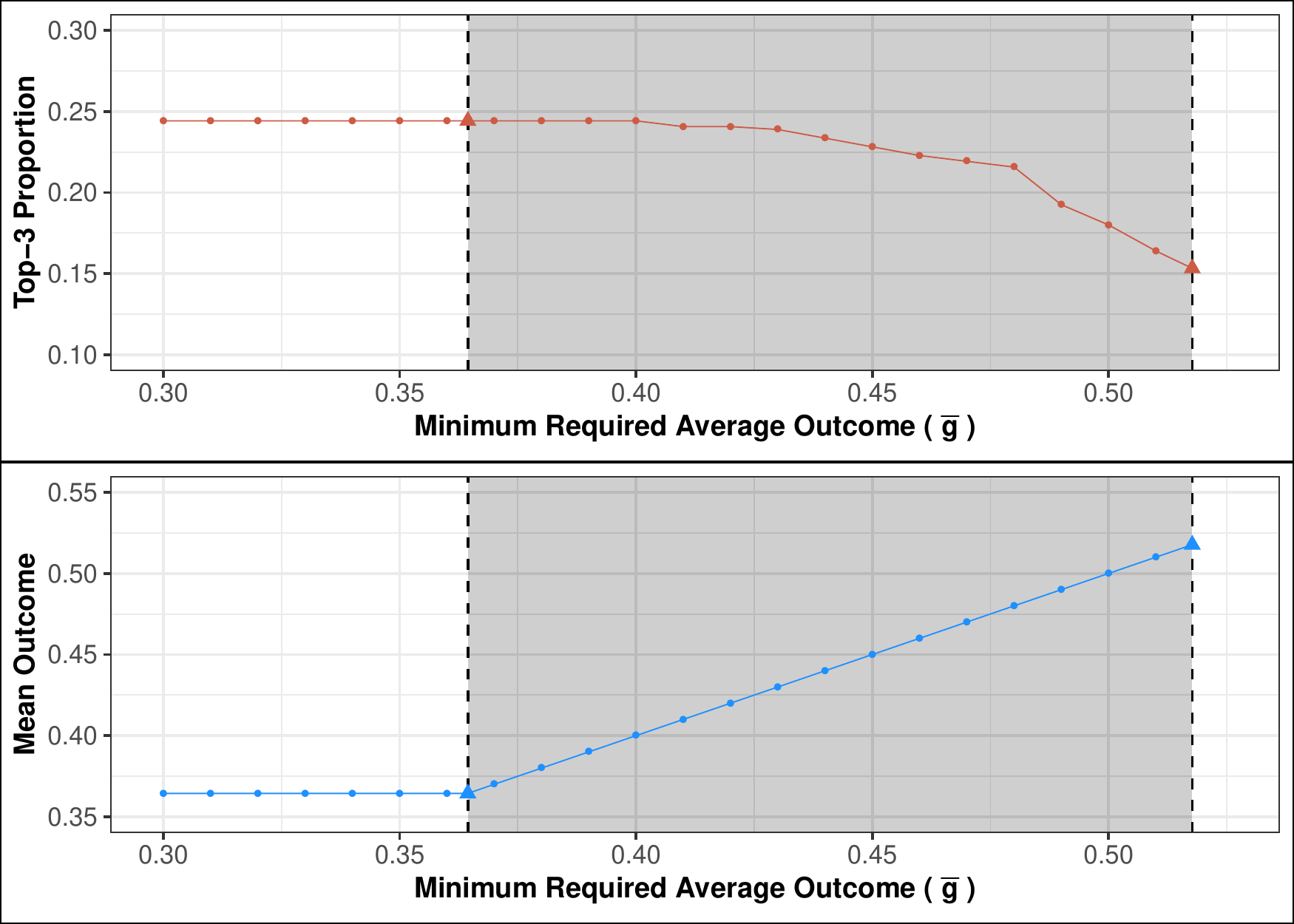}
\caption{Results of applying our $\bar g$-Constrained Priority Mechanism to refugee families in the United States (without truncated preferences) for various specified thresholds for the expected minimum level of average integration outcomes ($\bar g$). This figure shows the results of applying the mechanism to the same data as in the main text Figure 3, except that the families' preference rank vectors were not truncated in the application illustrated here. Upper panel shows the average probability that a refugee got assigned to one of their top three locations. Lower panel shows the realized average integration outcomes, i.e. the average projected probability of employment. $N=561$ families who arrived in Q3 of 2016.} \label{fig:lirs_CPMnotrunc}
\end{figure}

In applying our mechanism to the 2016 Q3 refugee data, we impose real-world assignment constraints, giving each location capacity for the same number of families as were sent to those locations in actuality. We also truncate each family's preference rank vector such that only the first 10 ranks are retained and indifference is established among the remaining locations. Figure 3 in the main text displays the results. In addition, Figure \ref{fig:lirs_CPMnotrunc} in this SI shows the results of the same simulations when the preference rank vectors are not truncated.

More details on the procedures used to generate the outcome score and preference rank vectors can be found below.

\subsection{Generating Outcome Scores and Preference Ranks}

The methods used for estimating the predicted probabilities of employment and outmigration in this study are the same as those employed in \cite{bansak2018improving}. The following material describes the procedures and is modified directly from the Supplementary Materials document of \cite{bansak2018improving}.

\subsection{Training vs. Prediction Data Designation}

Let $\textbf{T}$ (training data) be the matrix of refugee data, in which the unit of observation is a single refugee, that will be used for model training. The $\textbf{T}$ matrix contains the data for all working age refugees in our data who arrived starting in 2011 and up to (but not including) the third quarter of 2016. For each refugee we observe her assigned location, response variables of interest (employment for the outcome score and outmigration for the preference rank), and her full set of covariates.

Let $\textbf{R}$ (prediction data) be the matrix of data for the working age, free case refugees who arrived during the third quarter of 2016. This comprises the set of refugees to whom we applied our mechanism in this application. In a real-world application, these $\textbf{R}$ matrix data would correspond to new refugee arrivals and must include the same set of covariates as in the model training data. In contrast to the model training data, however, these prediction data need not include refugees' response variables. In fact, in a real-world prospective implementation of the mechanism, refugees belonging to these prediction data will not have yet been assigned to a resettlement location.

Note that when applying our mechanism both the model training and prediction data should be subsetted to the group of refugees for whom the outcomes of interest are relevant. In our application the integration outcome is employment and therefore the population of interest is working-age refugees. In addition, the prediction data should be subsetted only to those refugees who are free to be assigned to different resettlement locations---in contrast to refugees with predetermined geographic destinations due to family ties and other special circumstances---as this is the subset for whom the mechanism is designed to help with the assignment process. That said, the model training data need not be restricted to only free cases. Free-case and non-free-case refugees might be sufficiently dissimilar that forecasting free-case refugees' outcomes with models built using non-free-case data may seem problematic. This issue is addressed, however, by including case type as a predictor variable in the model building process (see below).

\subsection{Modeling}

The training data is used to build a bundle of learners that predict refugees' probabilities of the response variables (employment and outmigration), and those learned models are then applied to the prediction data to generate their predicted probabilities.

The modeling is implemented on a location-by-location basis. For each resettlement location, the training data are first subsetted to those refugees who were assigned to that location, and a statistical model is then fit that uses those refugees' characteristics to predict the response. That fitted model is then applied to the prediction data (2016 Q3 refugees) to predict the probability of the response for these refugee arrivals if they were hypothetically sent to the location in question. This process is performed separately for each individual location, which yields for each refugee in the prediction data a vector of predicted probabilities, one for each location. Collectively for all refugees in the prediction data, the final result is then a matrix of predicted probabilities ($\textbf{M}$ matrix) with rows representing individual refugees and columns representing resettlement locations. Note that there are two $\textbf{M}$ matrices: one for probabilities of employment and one for probabilities of outmigration.

More formally, for each refugee $r = 1, ..., n_T$, let the response of interest (e.g. employment) be denoted by $y_r \in \{ 0,1 \}$ and the location assignment denoted by $w_r \in \{1,...,k \}$, for a total of $k$ possible resettlement locations. Let $\vec{x}_r$ denote a $p$-dimensional feature vector comprised of the characteristics of refugee $r$, and $x_{rm}$ denote the $m$th feature in $\vec{x}_r$, where $m = 1, ..., p$. The goal of the modeling process is to learn the function $\theta_l(\vec{x}_r) = P(y_r = 1 | \vec{x}_r, w_r = l)$. The following describes the steps in the modeling stage.

\begin{enumerate}

\item Designate the historical model training data and denote it by the matrix $\textbf{T}$:

\[ \textbf{T} =
\left[   \begin{array}{ccccccc}
  y_1 & w_1 & x_{11} & \cdots & x_{1m} & \cdots & x_{1p} \\
  \vdots & \vdots & \vdots &  & \vdots &  & \vdots \\
  y_r & w_r & x_{r1} & \cdots & x_{rm} & \cdots & x_{rp} \\
  \vdots & \vdots & \vdots &  & \vdots &  & \vdots \\
  y_{n_T} & w_{n_T} & x_{n_T 1} & \cdots & x_{n_T m} & \cdots & x_{n_T p} \\
  \end{array} \right]
\]

\item Train a set of $k$ models, $\pmb{\Theta} = \{ \hat{\theta}_1(\vec{x}_r), ..., \hat{\theta}_l(\vec{x}_r), ..., \hat{\theta}_k(\vec{x}_r) \}$ as follows. 
\\ For $l = 1, ..., k$:

\begin{enumerate}

\item Subset $\textbf{T}$ to refugees for whom $w_r = l$ (i.e. refugees assigned to $l$-th location), and call this $\textbf{T}_l$:

\[ \textbf{T}_l =
\left[   \begin{array}{cccccc}
  y_1 & x_{11} & \cdots & x_{1m} & \cdots & x_{1p} \\
  \vdots & \vdots &  & \vdots &  & \vdots \\
  y_r & x_{r1} & \cdots & x_{rm} & \cdots & x_{rp} \\
  \vdots & \vdots &  & \vdots &  & \vdots \\
  y_{n_l} & x_{n_l1} & \cdots & x_{n_lm} & \cdots & x_{n_lp} \\
  \end{array} \right]_{w = l}
  =
  \left[   \begin{array}{cc}
  y_1 & \vec{x}_1 \\
  \vdots & \vdots  \\
  y_r & \vec{x}_r \\
  \vdots & \vdots \\
  y_{n_l} & \vec{x}_{n_l} \\
  \end{array} \right]_{w = l}
\]

where $n_l$ denotes the number of refugees for whom $w_r = l$.

\item Using the data in $\textbf{T}_l$ (the outcome $y_r$ and feature vector $\vec{x}_r$ for all $n_l$ refugees in $\textbf{T}_l$), model and estimate the function $\hat{\theta}_l(\vec{x}_r)$. 

\end{enumerate}

\item Designate the data on new refugee arrivals and denote them by the matrix $\textbf{R}$:

\[ \textbf{R} =
\left[   \begin{array}{ccccc}
  \dot{x}_{11} & \cdots & \dot{x}_{1m} & \cdots & \dot{x}_{1p} \\
  \vdots &  & \vdots &  & \vdots \\
  \dot{x}_{r1} & \cdots & \dot{x}_{rm} & \cdots & \dot{x}_{rp} \\
  \vdots &  & \vdots &  & \vdots \\
  \dot{x}_{n_R1} & \cdots & \dot{x}_{n_Rm} & \cdots & \dot{x}_{n_Rp} \\
  \end{array} \right]
  =
  \left[   \begin{array}{c}
  \vec{\dot{x}}_1 \\
  \vdots  \\
  \vec{\dot{x}}_r \\
  \vdots \\
  \vec{\dot{x}}_{n_R} \\
  \end{array} \right]
\]

where $n_R$ denotes the number of new refugee arrivals. 

The matrix $\textbf{R}$ corresponds to the 2016 Q3 refugees in this application.

\item For all refugees in $\textbf{R}$ and all resettlement locations, estimate $P(\dot{y}_r = 1 | \vec{\dot{x}}_r, \dot{w}_r = l)$ as follows. \\
For $r = 1, ..., n_R$:
\begin{itemize}
\item[] For $l = 1, ..., k$:

\begin{itemize}
\item[] Estimate $P(\dot{y}_r = 1 | \vec{\dot{x}}_r, \dot{w}_r = l)$ by applying $l$th model in $\pmb{\Theta}$ to $\vec{\dot{x}}_r$:
\item[] $\widehat{P}(\dot{y}_r = 1 | \vec{\dot{x}}_r, \dot{w}_r = l) = \hat{\theta}_l(\vec{\dot{x}}_r) \equiv \pi_{rl}$
\end{itemize}
\item[] Arrange the $\pi_{rl}$ into a vector, $\vec{\pi}_r = [\pi_{r1}, ..., \pi_{rk}]$.

\end{itemize}

\item Produce a matrix of predicted probabilities, with rows corresponding to new refugees and columns corresponding to resettlement locations, as follows. \\
Arrange vectors $\vec{\pi}_r$ into rows of the matrix $\textbf{M}$:

\[ \textbf{M} =
  \left[   \begin{array}{c}
  \vec{\pi}_1 \\
  \vdots  \\
  \vec{\pi}_r \\
  \vdots \\
  \vec{\pi}_{n_R} \\
  \end{array} \right]
  =
\left[   \begin{array}{ccccc}
  \pi_{11} & \cdots & \pi_{1l} & \cdots & \pi_{1k} \\
  \vdots &  & \vdots &  & \vdots \\
  \pi_{r1} & \cdots & \pi_{rl} & \cdots & \pi_{rk} \\
  \vdots &  & \vdots &  & \vdots \\
  \pi_{n_R1} & \cdots & \pi_{n_Rl} & \cdots & \pi_{n_Rk} \\
  \end{array} \right]
\]
This is the final modeling stage output.

\end{enumerate}

We follow \cite{bansak2018improving} and use boosted trees \citep{friedman2009elements, friedman2001greedy} to estimate $\hat{\theta}_l(\vec{x}_r)$ in step 2(b). See \cite{bansak2018improving} for more details on the selection criteria and model performance metrics leading to the choice of boosted trees. Specifically, we use stochastic gradient boosted trees (bag fraction of 0.5) with a binomial deviance loss function \citep{friedman2002stochastic, friedman2009elements}, which we implemented in \texttt{R} using the \texttt{gbm} package \citep{ridgeway2017generalized}.  Tuning parameter values, including the interaction depth, learning rate, and number of boosting iterations (the early stopping point) are selected via cross-validation within the training data for each location-specific model.

We use the following predictors: \emph{Free case}, \emph{Speaks English},  \emph{Age at arrival}, \emph{Male}, \emph{Education} (ordered variable differentiating between no/unknown education, less than secondary, secondary, technical/professional, and university),  \emph{Country of origin} (one binary variable for each of the largest origin groups including Burma, Iraq, Bhutan, Somalia, Afghanistan, Democratic Republic of Congo, Iran, Eritrea, Ukraine, Syria, Sudan, Ethiopia, and Moldova), \emph{Year of arrival}, and \emph{Month of arrival}.

\subsection{Mapping to Case-Level}

Since the assignment of refugees typically takes place at the level of the case (typically a family), we need to map the refugee-level predicted probabilities from the modeling process to a case-level metric. For each case-location pair, we apply the mapping function to the refugee-location predicted probabilities for all refugees belonging to that case, yielding a single value for that case-location pair. This results in a new matrix ($\textbf{M}^*$ matrix) with the same number of columns (locations) as previously but now as many rows as cases rather than refugees.

Formally, let $i = 1,...,n$ denote the refugee case, with a total of $n$ cases, where $n \leq n_R$. The mapping process then proceeds as follows:

\begin{enumerate}
\item Perform mapping of individual predicted probabilities to case-level metric as follows. \\ 
For $i = 1, ..., n$:
\begin{itemize}
\item[] For $l = 1,...,k$:
\begin{itemize}
\item[] Let $\tilde{\pi}_{il} = \{ \pi_{rl} \:\:\: \forall \:\:\: r \in i \}$. (That is, $\tilde{\pi}_{il}$ is the set of all $\pi_{rl}$ for the $l$th location and refugees belonging to the $i$th case.)
\item[] Compute $\gamma_{il} = \psi(\tilde{\pi}_{il})$ where $\psi$ is a predetermined mapping function.
\end{itemize}
\item[] Arrange the $\gamma_{il}$ into a vector, $\vec{\gamma}_i = [\gamma_{i1}, ..., \gamma_{ik}]$.
\end{itemize}
\item Produce a matrix containing the case-level metric for all case-location pairs, with rows corresponding to cases and columns corresponding to resettlement locations, as follows. \\
Arrange vectors $\vec{\gamma}_i$ produced in step 1 into rows of the matrix $\textbf{M}^*$:

\[ \textbf{M}^* =
  \left[   \begin{array}{c}
  \vec{\gamma}_1 \\
  \vdots  \\
  \vec{\gamma}_i \\
  \vdots \\
  \vec{\gamma}_{n} \\
  \end{array} \right]
  =
\left[   \begin{array}{ccccc}
  \gamma_{11} & \cdots & \gamma_{1l} & \cdots & \gamma_{1k} \\
  \vdots &  & \vdots &  & \vdots \\
  \gamma_{i1} & \cdots & \gamma_{il} & \cdots & \gamma_{ik} \\
  \vdots &  & \vdots &  & \vdots \\
  \gamma_{n1} & \cdots & \gamma_{nl} & \cdots & \gamma_{nk} \\
  \end{array} \right]
\]
This is the final mapping stage output.
\end{enumerate}

In step 1, the function $\psi$ must be specified. In our application, we employ the mean for both the predicted probabilities of employment and the predicted probabilities of outmigration (see \cite{bansak2018improving} for alternative choices).

\subsection{Final Construction of Outcome Scores and Preference Ranks}

The $\textbf{M}^*$ matrix pertaining to the predicted probabilities of employment directly provides the outcome scores for use in the mechanism. However, the $\textbf{M}^*$ matrix pertaining to the predicted probabilities of outmigration must be further transformed to provide the (inferred) preference ranks. Specifically, for each row (case), we rank locations such that the location with the lowest (highest) average probability of outmigration is ranked first (last), producing a preference rank vector for each case.

\subsection{Alternative Method for Estimating Preferences via Structural Adjustment}

We also examined an alternative method to estimate location preferences using a model that explicitly corrects for potential bias due to relocation costs. Note that in our data we only observe whether a refugee out-migrates from her initial resettlement location or not. This decision will be a function of both location preferences (i.e. outmigration should be higher in less desirable locations) and the costs of relocation (e.g. varying geographic and economic factors may result in higher costs of relocating from certain locations). 

Here we leverage a structural model of outmigration to isolate the component of outmigration that is likely attributable to location preferences, rather than the costs of relocation. In particular, we follow standard models in the literature on immigrant and refugee location choices and estimate the following structural model of outmigration:
$$
y_{ijt}=\alpha + \beta X_{jt} + \theta_j + \phi_t + \theta_j\times t + \epsilon_{ijt} 
$$
where $y_{ijt}$ is the outcome of whether refugee $i$ who arrived in year $t$ out-migrates from her initial resettlement location $j$, $X_{jt}$ are a set of time-varying location specific characteristics (e.g. rental prices, unemployment rates, ethnic networks, welfare generosity, etc.) that affect the costs of relocation with coefficients $\beta$, $\theta_j $ is a set of location specific fixed effects that capture all time-invariant factors that affect the costs of relocation (e.g. remote location), $\phi_t$ is a set of year fixed effects that capture common shocks (e.g. changes in transportation costs), and $\theta_j \times t$ is a set of location-specific linear time trends that capture changes in location-specific relocation costs that have a linear effect on outmigration (e.g. local economic decline, changes in local transportation infrastructure, etc.). 

We include in $X$ a set of location characteristics that are commonly included in structural models of location choices \citep{borjas1999immigration,zavodny1999determinants,damm2009determinants,aaslund2007and,mossad2019search}. In particular, we include the local unemployment rate and personal income per capita to proxy for economic opportunities \citep{aaslund2007and,damm2009determinants,mossad2019search}, rental prices to proxy for cost of living \citep{damm2009determinants,mossad2019search}, ethnic shares to proxy for enclave effects \citep{beaman2012social,mossad2019search}, and welfare spending per capita to proxy for welfare magnet effects \citep{damm2009determinants,borjas1999immigration}. A list of definitions and sources are provided below. To merge in this information, we first identified the county of each resettlement location and then merged in the location-specific characteristics measured at the refugee's time of arrival.

We fit the model with a logistic link function on the training data of refugees that arrived prior to the third quarter of 2016. Note that to fit this model we restrict the training data to only refugees who arrived as free cases. Since free cases do not choose their initial resettlement location but are exogenously placed by the resettlement agencies, this sample restriction limits potential bias due to individuals sorting into initial locations based on unobserved characteristics such as location preferences (see \cite{aaslund2007and} for a similar identification that leverages a placement policy in Sweden to estimate location preferences). 

We then use the fitted model to generate the predicted probabilities for outmigration for each family in the test set of quarter 3 2016 arrivals for each resettlement location. These predictions capture the probabilities of outmigration that we would expect for a given family purely based on the location specific relocation costs as captured by the structural model. 

In the next step, we then compute the difference between the predicted probabilities from our previous model that was based on individual-level characteristics and the predicted probabilities from the structural model. The resulting differences can then be interpreted as variation in outmigration that is mostly driven by location preferences because it is adjusted for the variation in outmigration that is driven by structural relocation costs. 

For example, consider a family who has a very low predicted probability of outmigration in a given location based on the individual model, but based on the structural model the predicted probability of outmigration in the same location is very high. This would indicate that this family has a strong preference to remain in this location even though based on the structural factors they would be pulled towards relocating. On the flip side, a family that has a very high predicted probability of outmigration based on the individual model but a very low probability of outmigration based on the structural model would suggest that they have a strong preference against living in this location given that the structural factors would pull them towards staying. 

Accordingly, as a last step for each family, we rank the locations based on the differences in predicted probabilities such that the location with the most negative (most positive) difference is ranked as most (least) preferred. 

The list of geographic factors and data sources is as follows:
\begin{itemize}
    \item \emph{Annual unemployment rate} in county. Data retrieved from the Local Area Unemployment Statistics (LAUS) from the Bureau of Labor Statistics. 
    \item \emph{Monthly Rental Price Index} in county. Data retrieved from Zillow Rent Index (ZRI) (Time Series Multifamily, SFR, and Condo/Coops). The Zillow Rent Index is a smoothed measure of the typical estimated market rate rent across a given region and housing type. We linearly interpolated for missing values.
    \item \emph{Personal annual income per capita} in county. Data retrieved from the Bureau of Economic Analysis.
    \item \emph{Share of co-nationals} in metro area. Share of co-nationals in each county year is estimated based on ACS 5 year and 3 year samples (downloaded from IPUMs, using the BLP and MET2013 variables). For some resettlement locations that were outside a metro area we merge based on city or PUMA instead of metro area. 
    \item \emph{Total annual state and local welfare spending per capita}. Data retrieved from Annual Survey of State and Local Government Finances US Census. 
\end{itemize}

\begin{figure}[tbhp]
\centering
\includegraphics[width=1\linewidth]{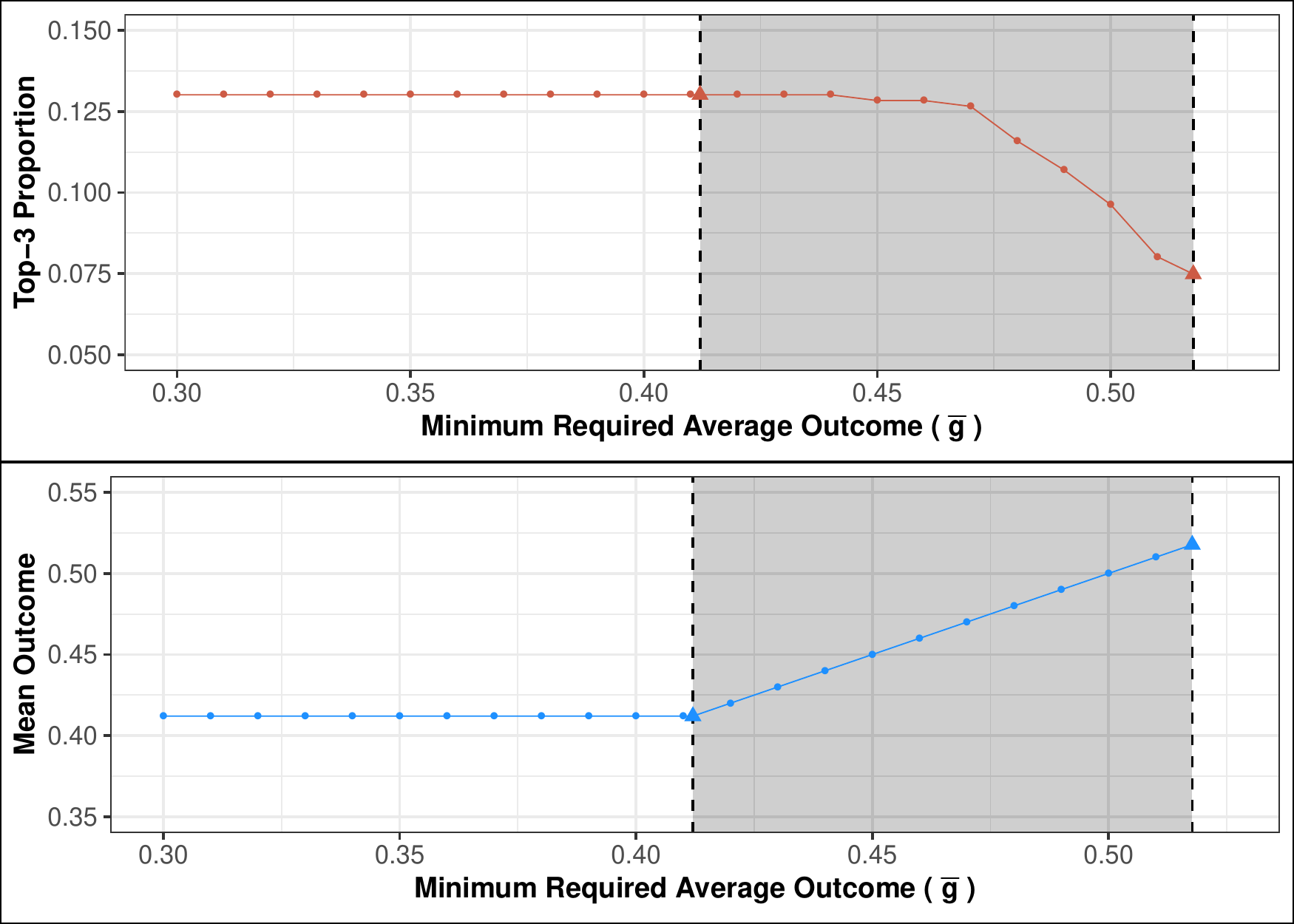}
\caption{Results of applying our $\bar g$-Constrained Priority Mechanism to refugee families in the United States for various specified thresholds for the expected minimum level of average integration outcomes ($\bar g$), with structurally adjusted preference estimates. This figure shows the results of applying the mechanism to the same data as in the main text Figure 3, except that the families' preference rank vectors were adjusted using a structural model design to account for differential relocation costs across locations. Upper panel shows the average probability that a refugee got assigned to one of their top three locations. Lower panel shows the realized average integration outcomes, i.e. the average projected probability of employment. $N=561$ families who arrived in Q3 of 2016.} \label{fig:lirs_CPMstructual}
\end{figure}

The results of applying our mechanism to the 2016 Q3 refugee data using these new preference estimates is shown in Figure \ref{fig:lirs_CPMstructual}.

\section{Education Application}

\subsection{Background Information on Application}

Here we illustrate our mechanism by applying them to a hypothetical example of choice of elementary schools. We consider a case where a school district might be interested to assign incoming Kindergarten students to elementary schools in the district with the goal to maximize academic achievement as measured by scores on standardized tests that are administered at the end of the Kindergarten grade. Students have preferences over schools and so the goal of the mechanism is to optimize on test scores and preferences subject to the minimum expected average level of test score set by the district.   

\subsection{Tennesse Star Data}

We leverage data from the Tennessee's Student Teacher Achievement Ratio (STAR) project conducted by the Tennessee State Department of Education. This data include student level data on from a longitudinal experiment in Tennessee that began in 1985 and tracked a cohort of students progressing from kindergarten through third grade (for details on the data and sample see \citet{SIWH9F_2008}). The data includes demographic information on the students, indicators for the schools that they attended, as well as information on achievement tests that were administered annually at the end of each grade. We focus on the sample of 1,674 students from 33 schools that are observed for all grades from Kindergarten through 3rd grade and have non-missing data for tests scores and background characteristics.

Table \ref{tab:destatEdu} shows the descriptive statistics for our sample. Below is a list of variables and measures used:
\begin{itemize}
\item \emph{Month of birth}: This variable is coded with values from 1 to 12.
\item \emph{Year of birth}: This variable is coded with values including 1978, 1979, 1980, and 1981
\item \emph{Race}: The student's race coded as six categories including White, Black, Asian, Hispanic, Native American, and Other.
\item \emph{Free lunch}: Binary variable coded as 1 if the student was eligible for free/reduced lunch in Kindergarten  and zero otherwise.
\item \emph{Special Education}: Binary variable coded as 1 if the student was eligible for special education status in Kindergarten and zero otherwise.
\item \emph{Female}: Binary variable coded as 1 for female students and zero otherwise
\item \emph{SAT Score Reading}: Total reading scaled score on the Stanford Achievement Test at the end of Kindergarten.
\item \emph{SAT Score Math}: Total math scaled score on the Stanford Achievement Test at the end of Kindergarten.
\item \emph{SAT Score Listening}: Total listening scaled score on the Stanford Achievement Test at the end of Kindergarten.
\item \emph{Sum of SAT Scores}: Sum of the three SAT scores for Reading, Math, and Total listening scaled score at the end of Kindergarten.
\item \emph{Left Kindergarten}: Variable used to measure outmigration from the Kindergarten school. Higher values indicate that the student left the Kindergarten school faster which can be interpreted as a stronger preference for another school. Coded 0 if student remained in the Kindergarten school for 1st, 2nd, and 3rd grade; coded 1 if student stay in Kindergarten school for grade 1 and 2 but left for another school for grade 3; coded 2 if student stay in Kindergarten school for grade 1, but left for another school for grade 2; and coded 3 if students left for another school for grade 1.
\end{itemize}

\subsection{Applying the Mechanism}

To generate each student's outcome score vector across each of the schools, we used the same stochastic gradient boosted tree models as in the refugee application to predict the expected tests score of a student at any of the schools, as a function of their background characteristics. The background characteristics included the students' age, gender, race, as well as information on whether they are eligible for free school lunches (a proxy for socioeconomic status) or special education. The test score outcome was defined as the sum of reading, math, and listening scaled SAT scores for the Kindergarten level. Given the small sample size for some schools we used the same data for the training and validation set and increased the bag fraction to 1. We look for the best fitting tree models over interactions depth of 3 to 8 using 5-fold cross-validation with total of 1,500 trees.

To generate the school preferences we inferred revealed school preferences of students from the observed transfers out of the schools. Specifically, we used the same modeling procedure of stochastic gradient boosted tree model as for the test scores but instead used a response variable that measured whether a student had transferred to another school by the first, second, or third grade. Based on these models we can then predict for each student the propensity for leaving each school as a function of their background characteristics. For each student, we then rank schools such that the school with the lowest (highest) propensity for transferring out is ranked first (last). In contrast to the refugee application there is no mapping to a case level since assignments are done at the student level. We impose the constraint that every school can only receive as many students as the did in actuality.

\clearpage

\section{Tables}

\begin{table}[!hbt]
  \centering
  \small
  \caption{Descriptive Statistics for United States Refugee Sample}\label{tab:destatUS}
    \begin{tabular}{lcc}
    \hline \hline
          &    Mean & SD     \\
\hline
 Male &      0.53 &      0.50 \\
 Speaks English &      0.42 &      0.49 \\
Age: \\
$\,\,$ 18-29 &      0.44 &      0.50 \\
$\,\,$ 30-39 &      0.28 &      0.45 \\
$\,\,$ 40-49 &      0.16 &      0.37 \\
$\,\,$ 50+ &      0.11 &      0.31 \\
Education: \\
$\,\,$ None/Unknown &      0.18 &      0.39 \\
$\,\,$ Less than Secondary &      0.39 &      0.49 \\
$\,\,$ Secondary &      0.21 &      0.41 \\
$\,\,$ Advanced &      0.10 &      0.30 \\
$\,\,$ University &      0.12 &      0.33 \\
Origin: \\
$\,\,$ Burma &      0.23 &      0.42 \\
$\,\,$ Iraq &      0.20 &      0.40 \\
$\,\,$ Bhutan &      0.13 &      0.34 \\
$\,\,$ Somalia &      0.11 &      0.31 \\
$\,\,$ Afghanistan &      0.07 &      0.25 \\
$\,\,$ Other &      0.26 &      0.44 \\
 Employed &      0.23 &      0.42 \\
    \hline \hline
\multicolumn{3}{p{0.45\textwidth}}{\scriptsize Sample consists of refugees of working age that were resettled by one of the largest resettlement agencies and arrived in the period from quarter 1, 2011 to quarter 3, 2016. N = 33,782.}
    \end{tabular} 
 \end{table}

\clearpage

\begin{table}[!hbt]
  \centering
  \small
  \caption{Descriptive Statistics for Student Sample}\label{tab:destatEdu}
    \begin{tabular}{lcc}
    \hline \hline
          &    Mean & SD     \\
\hline
Month of Birth &    6.22 &  3.45 \\ 
Year of Birth & 1979.74 &  0.45 \\ 
Race: \\
  White &    0.82 &  0.38 \\ 
  Black &    0.17 &  0.38 \\ 
  Asian &    0.00 &  0.05 \\ 
  Hispanic &    0.00 &  0.02 \\ 
  Native American &    0.00 &  0.00 \\ 
  Other &    0.00 &  0.03 \\ 

  Free Lunch &    0.34 &  0.47 \\ 
  Special Education &    0.02 &  0.13 \\ 
  Female &    0.51 &  0.50 \\ 
  SAT Score Reading &  445.34 & 31.17 \\ 
  SAT Score Math &  499.43 & 43.41 \\ 
  SAT Score Listening &  547.18 & 30.38 \\ 
  Sum of SAT Scores & 1491.95 & 88.22 \\ 
  Left Kindergarten &    0.06 &  0.37 \\ 
    \hline \hline
\multicolumn{3}{p{0.45\textwidth}}{\scriptsize Sample consists of students from the Tennessee Star data. N = 1,674. Note that the ``Left Kindergarten" variable denotes the number of years during K-3 that a student was in a school that was different from their Kindergarten school.}
    \end{tabular} 
 \end{table} 

\clearpage

\section{Figures}

\clearpage

\clearpage

\begin{figure}
\centering
\includegraphics[width=1\linewidth]{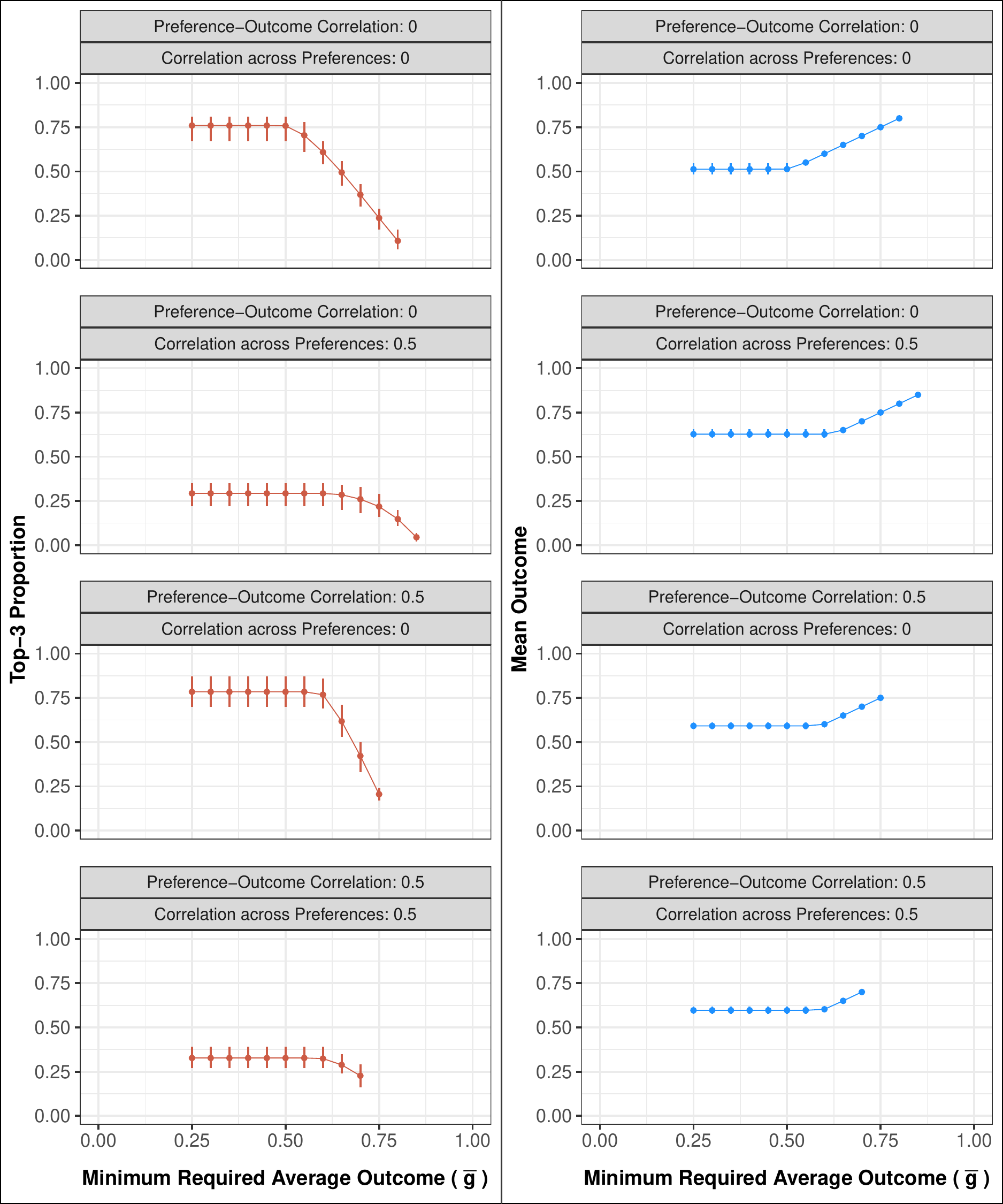}
\caption{Results from re-running simulations discussed in the main text Simulation Data Section and shown in Figure 1, where at each level of $\bar g$, the mechanism is applied $100$ separate times and the order of the agents is re-randomized each time. The dots denote the results---the proportion assigned to a top-$3$ location in the panels on the left, and the mean outcome score on the right---averaged across the $100$ re-orderings, and the intervals denote the maximum and minimum results obtained across the $100$ re-orderings.}
\label{fig:reordering}
\end{figure}

\clearpage

\bibliography{references}

\begin{thebibliography}{48}
\newcommand{\enquote}[1]{``#1''}
\expandafter\ifx\csname natexlab\endcsname\relax\def\natexlab#1{#1}\fi

\bibitem[\protect\citeauthoryear{Abdulkadiro{\u{g}}lu, Pathak, and
  Roth}{Abdulkadiro{\u{g}}lu et~al.}{2009}]{abdulkadirouglu2009strategy}
\textsc{Abdulkadiro{\u{g}}lu, A., P.~A. Pathak, and A.~E. Roth} (2009):
  \enquote{Strategy-proofness versus efficiency in matching with indifferences:
  Redesigning the NYC high school match,} \emph{American Economic Review}, 99,
  1954--78.

\bibitem[\protect\citeauthoryear{Abdulkadiroglu and Sonmez}{Abdulkadiroglu and
  Sonmez}{1998}]{abdulkadiroglu1998random}
\textsc{Abdulkadiroglu, A. and T.~Sonmez} (1998): \enquote{Random serial
  dictatorship and the core from random endowments in house allocation
  problems,} \emph{Econometrica}, 66, 689.

\bibitem[\protect\citeauthoryear{Abdulkadiro{\u{g}}lu and
  S{\"o}nmez}{Abdulkadiro{\u{g}}lu and
  S{\"o}nmez}{2003}]{abdulkadirouglu2003school}
\textsc{Abdulkadiro{\u{g}}lu, A. and T.~S{\"o}nmez} (2003): \enquote{School
  choice: A mechanism design approach,} \emph{American Economic Review}, 93,
  729--747.

\bibitem[\protect\citeauthoryear{Abdulkadiroglu and S{\"o}nmez}{Abdulkadiroglu
  and S{\"o}nmez}{2013}]{abdulkadiroglu2013matching}
\textsc{Abdulkadiroglu, A. and T.~S{\"o}nmez} (2013): \enquote{Matching
  markets: Theory and practice,} \emph{Advances in Economics and Econometrics},
  1, 3--47.

\bibitem[\protect\citeauthoryear{Acharya, Bansak, and Hainmueller}{Acharya
  et~al.}{2020{\natexlab{a}}}]{codeocean2020abh}
\textsc{Acharya, A., K.~Bansak, and J.~Hainmueller} (2020{\natexlab{a}}):
  \enquote{{Replication Materials for: Combining Outcome-Based and
  Preference-Based Matching: A Constrained Priority Mechanism},}
  \url{https://doi.org/10.24433/CO.3735899.v1}, Code Ocean, V1.

\bibitem[\protect\citeauthoryear{Acharya, Bansak, and Hainmueller}{Acharya
  et~al.}{2020{\natexlab{b}}}]{DVN/ZEV0WX}
---\hspace{-.1pt}---\hspace{-.1pt}--- (2020{\natexlab{b}}):
  \enquote{{Replication Materials for: Combining Outcome-Based and
  Preference-Based Matching: A Constrained Priority Mechanism},}
  \url{https://doi.org/10.7910/DVN/ZEV0WX}, Harvard Dataverse, V1.

\bibitem[\protect\citeauthoryear{Achilles, Bain, Bellott, Boyd-Zaharias, Finn,
  Folger, Johnston, and Word}{Achilles et~al.}{2008}]{SIWH9F_2008}
\textsc{Achilles, C., H.~P. Bain, F.~Bellott, J.~Boyd-Zaharias, J.~Finn,
  J.~Folger, J.~Johnston, and E.~Word} (2008): \enquote{{Tennessee's Student
  Teacher Achievement Ratio (STAR) project},}
  Https://doi.org/10.7910/DVN/SIWH9F, Harvard Dataverse.

\bibitem[\protect\citeauthoryear{Andersson and Ehlers}{Andersson and
  Ehlers}{2016}]{andersson2016assigning}
\textsc{Andersson, T. and L.~Ehlers} (2016): \enquote{Assigning refugees to
  landlords in Sweden: Stable maximum matchings,} Tech. rep., Working Paper,
  Lund University.

\bibitem[\protect\citeauthoryear{{\AA}slund and Rooth}{{\AA}slund and
  Rooth}{2007}]{aaslund2007and}
\textsc{{\AA}slund, O. and D.-O. Rooth} (2007): \enquote{Do when and where
  matter? Initial labour market conditions and immigrant earnings,} \emph{The
  Economic Journal}, 117, 422--448.

\bibitem[\protect\citeauthoryear{Bansak}{Bansak}{2020}]{bansak2020minimumrisk}
\textsc{Bansak, K.} (2020): \enquote{A minimum-risk dynamic assignment
  mechanism along with an approximation, heuristics, and extension from single
  to batch assignments,} \emph{arXiv preprint arXiv:2007.03069}.

\bibitem[\protect\citeauthoryear{Bansak, Ferwerda, Hainmueller, Dillon,
  Hangartner, Lawrence, and Weinstein}{Bansak
  et~al.}{2018}]{bansak2018improving}
\textsc{Bansak, K., J.~Ferwerda, J.~Hainmueller, A.~Dillon, D.~Hangartner,
  D.~Lawrence, and J.~Weinstein} (2018): \enquote{Improving refugee integration
  through data-driven algorithmic assignment,} \emph{Science}, 359, 325--329.

\bibitem[\protect\citeauthoryear{Beaman}{Beaman}{2012}]{beaman2012social}
\textsc{Beaman, L.~A.} (2012): \enquote{Social networks and the dynamics of
  labour market outcomes: Evidence from refugees resettled in the US,}
  \emph{The Review of Economic Studies}, 79, 128--161.

\bibitem[\protect\citeauthoryear{Bertsekas and Tseng}{Bertsekas and
  Tseng}{1994}]{bertsekas1994relax}
\textsc{Bertsekas, D.~P. and P.~Tseng} (1994): \enquote{RELAX-IV: {A} faster
  version of the RELAX code for solving minimum cost flow problems,} Tech.
  rep., Massachusetts Institute of Technology, Laboratory for Information and
  Decision Systems Cambridge, MA.

\bibitem[\protect\citeauthoryear{Borjas}{Borjas}{1999}]{borjas1999immigration}
\textsc{Borjas, G.~J.} (1999): \enquote{Immigration and welfare magnets,}
  \emph{Journal of Labor Economics}, 17, 607--637.

\bibitem[\protect\citeauthoryear{Burkard, Dell'Amico, and Martello}{Burkard
  et~al.}{2012}]{burkard2012assignmentch4}
\textsc{Burkard, R., M.~Dell'Amico, and S.~Martello} (2012): \emph{Assignment
  Problems}, Society for Industrial and Applied Mathematics.

\bibitem[\protect\citeauthoryear{Damm}{Damm}{2009}]{damm2009determinants}
\textsc{Damm, A.~P.} (2009): \enquote{Determinants of recent immigrants’
  location choices: {Quasi-experimental} evidence,} \emph{Journal of Population
  Economics}, 22, 145--174.

\bibitem[\protect\citeauthoryear{Damm}{Damm}{2014}]{damm2014neighborhood}
---\hspace{-.1pt}---\hspace{-.1pt}--- (2014): \enquote{Neighborhood quality and
  labor market outcomes: Evidence from quasi-random neighborhood assignment of
  immigrants,} \emph{Journal of Urban Economics}, 79, 139--166.

\bibitem[\protect\citeauthoryear{Delacr{\'e}taz, Kominers, and
  Teytelboym}{Delacr{\'e}taz et~al.}{2016}]{delacretaz2016refugee}
\textsc{Delacr{\'e}taz, D., S.~D. Kominers, and A.~Teytelboym} (2016):
  \enquote{Refugee resettlement,} Tech. rep., Working Paper, University of
  Melbourne.

\bibitem[\protect\citeauthoryear{Dur, Kominers, Pathak, and S{\"o}nmez}{Dur
  et~al.}{2018}]{dur2018reserve}
\textsc{Dur, U., S.~D. Kominers, P.~A. Pathak, and T.~S{\"o}nmez} (2018):
  \enquote{Reserve design: Unintended consequences and the demise of Boston's
  walk zones,} \emph{Journal of Political Economy}, 126, 2457--2479.

\bibitem[\protect\citeauthoryear{Echenique and Yenmez}{Echenique and
  Yenmez}{2015}]{echenique2015control}
\textsc{Echenique, F. and M.~B. Yenmez} (2015): \enquote{How to control
  controlled school choice,} \emph{American Economic Review}, 105, 2679--94.

\bibitem[\protect\citeauthoryear{Ehlers, Hafalir, Yenmez, and Yildirim}{Ehlers
  et~al.}{2014}]{ehlers2014school}
\textsc{Ehlers, L., I.~E. Hafalir, M.~B. Yenmez, and M.~A. Yildirim} (2014):
  \enquote{School choice with controlled choice constraints: Hard bounds versus
  soft bounds,} \emph{Journal of Economic Theory}, 153, 648--683.

\bibitem[\protect\citeauthoryear{Fern{\'a}ndez-Huertas~Moraga and
  Rapoport}{Fern{\'a}ndez-Huertas~Moraga and
  Rapoport}{2015}]{fernandez2015tradable}
\textsc{Fern{\'a}ndez-Huertas~Moraga, J. and H.~Rapoport} (2015):
  \enquote{Tradable refugee-admission quotas and EU asylum policy,}
  \emph{CESifo Economic Studies}, 61, 638--672.

\bibitem[\protect\citeauthoryear{Friedman}{Friedman}{2001}]{friedman2001greedy}
\textsc{Friedman, J.~H.} (2001): \enquote{Greedy function approximation: {A}
  gradient boosting machine,} \emph{Annals of Statistics}, 29, 1189--1232.

\bibitem[\protect\citeauthoryear{Friedman}{Friedman}{2002}]{friedman2002stochastic}
---\hspace{-.1pt}---\hspace{-.1pt}--- (2002): \enquote{Stochastic gradient
  boosting,} \emph{Computational Statistics \& Data Analysis}, 38, 367--378.

\bibitem[\protect\citeauthoryear{Friedman, Hastie, and Tibshirani}{Friedman
  et~al.}{2009}]{friedman2009elements}
\textsc{Friedman, J.~H., T.~Hastie, and R.~Tibshirani} (2009): \emph{The
  Elements of Statistical Learning, 2nd ed.}, Springer.

\bibitem[\protect\citeauthoryear{Gale and Shapley}{Gale and
  Shapley}{1962}]{gale1962college}
\textsc{Gale, D. and L.~S. Shapley} (1962): \enquote{College admissions and the
  stability of marriage,} \emph{The American Mathematical Monthly}, 69, 9--15.

\bibitem[\protect\citeauthoryear{G{\"o}lz and Procaccia}{G{\"o}lz and
  Procaccia}{2019}]{golz2019migration}
\textsc{G{\"o}lz, P. and A.~D. Procaccia} (2019): \enquote{Migration as
  submodular optimization,} in \emph{Proceedings of the AAAI Conference on
  Artificial Intelligence}, vol.~33, 549--556.

\bibitem[\protect\citeauthoryear{Hansen and Klopfer}{Hansen and
  Klopfer}{2006}]{hansen2006optimal}
\textsc{Hansen, B.~B. and S.~O. Klopfer} (2006): \enquote{Optimal full matching
  and related designs via network flows,} \emph{Journal of Computational and
  Graphical Statistics}, 15, 609--627.

\bibitem[\protect\citeauthoryear{Kamada and Kojima}{Kamada and
  Kojima}{2015}]{kamada2015efficient}
\textsc{Kamada, Y. and F.~Kojima} (2015): \enquote{Efficient matching under
  distributional constraints: Theory and applications,} \emph{American Economic
  Review}, 105, 67--99.

\bibitem[\protect\citeauthoryear{Kuhn}{Kuhn}{1955}]{kuhn1955hungarian}
\textsc{Kuhn, H.~W.} (1955): \enquote{The Hungarian method for the assignment
  problem,} \emph{Naval Research Logistics (NRL)}, 2, 83--97.

\bibitem[\protect\citeauthoryear{Lasswell}{Lasswell}{1936}]{laswell1936politics}
\textsc{Lasswell, H.~D.} (1936): \emph{Politics: Who Gets What, When, How},
  Cleveland: Meridian Books.

\bibitem[\protect\citeauthoryear{Mart{\'e}n, Hainmueller, and
  Hangartner}{Mart{\'e}n et~al.}{2019}]{marten2019ethnic}
\textsc{Mart{\'e}n, L., J.~Hainmueller, and D.~Hangartner} (2019):
  \enquote{Ethnic networks can foster the economic integration of refugees,}
  \emph{Proceedings of the National Academy of Sciences}, 116, 16280--16285.

\bibitem[\protect\citeauthoryear{Milgrom and Tadelis}{Milgrom and
  Tadelis}{2018}]{milgrom2018artificial}
\textsc{Milgrom, P.~R. and S.~Tadelis} (2018): \enquote{How Artificial
  Intelligence and Machine Learning Can Impact Market Design,} Tech. rep.,
  National Bureau of Economic Research.

\bibitem[\protect\citeauthoryear{Moraga and Rapoport}{Moraga and
  Rapoport}{2014}]{moraga2014tradable}
\textsc{Moraga, J. F.-H. and H.~Rapoport} (2014): \enquote{Tradable immigration
  quotas,} \emph{Journal of Public Economics}, 115, 94--108.

\bibitem[\protect\citeauthoryear{Mossad, Ferwerda, Lawrence, Weinstein, and
  Hainmueller}{Mossad et~al.}{2020}]{mossad2019search}
\textsc{Mossad, N., J.~Ferwerda, D.~Lawrence, J.~M. Weinstein, and
  J.~Hainmueller} (2020): \enquote{In search of opportunity and community:
  {The} secondary migration of refugees in the {United States},} \emph{Science
  Advances}, forthcoming.

\bibitem[\protect\citeauthoryear{Mousa}{Mousa}{2018}]{mousa2018boosting}
\textsc{Mousa, S.} (2018): \enquote{Boosting Refugee Outcomes: Evidence from
  Policy, Academia, and Social Innovation,} \emph{SSRN Working Paper}.

\bibitem[\protect\citeauthoryear{Munkres}{Munkres}{1957}]{munkres1957algorithms}
\textsc{Munkres, J.} (1957): \enquote{Algorithms for the assignment and
  transportation problems,} \emph{Journal of the Society for Industrial and
  Applied Mathematics}, 5, 32--38.

\bibitem[\protect\citeauthoryear{Narita}{Narita}{2019}]{narita2019experiment}
\textsc{Narita, Y.} (2019): \enquote{Experiment-as-Market: Incorporating
  Welfare into Randomized Controlled Trials,} \emph{Available at SSRN 3094905}.

\bibitem[\protect\citeauthoryear{Pathak}{Pathak}{2011}]{pathak2011mechanism}
\textsc{Pathak, P.~A.} (2011): \enquote{The mechanism design approach to
  student assignment,} \emph{Annu. Rev. Econ.}, 3, 513--536.

\bibitem[\protect\citeauthoryear{Pathak}{Pathak}{2016}]{pathak2016really}
---\hspace{-.1pt}---\hspace{-.1pt}--- (2016): \emph{What really matters in
  designing school choice mechanisms}.

\bibitem[\protect\citeauthoryear{Ridgeway}{Ridgeway}{2017}]{ridgeway2017generalized}
\textsc{Ridgeway, G.} (2017): \enquote{Package `gbm',} Tech. rep.,
  Comprehensive R Archive Network (CRAN),
  \url{https://cran.r-project.org/web/packages/gbm/gbm.pdf}.

\bibitem[\protect\citeauthoryear{Roth}{Roth}{1982}]{roth1982incentive}
\textsc{Roth, A.~E.} (1982): \enquote{Incentive compatibility in a market with
  indivisible goods,} \emph{Economics letters}, 9, 127--132.

\bibitem[\protect\citeauthoryear{Roth}{Roth}{2015}]{roth2015gets}
---\hspace{-.1pt}---\hspace{-.1pt}--- (2015): \emph{Who Gets What -- and Why:
  The New Economics of Matchmaking and Market Design}, Houghton Mifflin
  Harcourt.

\bibitem[\protect\citeauthoryear{Roth}{Roth}{2018}]{roth2018marketplaces}
---\hspace{-.1pt}---\hspace{-.1pt}--- (2018): \enquote{Marketplaces, markets,
  and market design,} \emph{American Economic Review}, 108, 1609--58.

\bibitem[\protect\citeauthoryear{Satterthwaite and Sonnenschein}{Satterthwaite
  and Sonnenschein}{1981}]{satterthwaite1981strategy}
\textsc{Satterthwaite, M.~A. and H.~Sonnenschein} (1981):
  \enquote{Strategy-proof allocation mechanisms at differentiable points,}
  \emph{The Review of Economic Studies}, 48, 587--597.

\bibitem[\protect\citeauthoryear{Shapley and Scarf}{Shapley and
  Scarf}{1974}]{shapley1974cores}
\textsc{Shapley, L. and H.~Scarf} (1974): \enquote{On cores and
  indivisibility,} \emph{Journal of mathematical economics}, 1, 23--37.

\bibitem[\protect\citeauthoryear{Trapp, Teytelboym, Martinello, Andersson, and
  Ahani}{Trapp et~al.}{2018}]{trapp2018placement}
\textsc{Trapp, A.~C., A.~Teytelboym, A.~Martinello, T.~Andersson, and N.~Ahani}
  (2018): \enquote{Placement Optimization in Refugee Resettlement,} Tech. rep.,
  Working Paper, Lund University.

\bibitem[\protect\citeauthoryear{Zavodny}{Zavodny}{1999}]{zavodny1999determinants}
\textsc{Zavodny, M.} (1999): \enquote{Determinants of recent immigrants’
  locational choices,} \emph{International Migration Review}, 33, 1014--1030.

\end{thebibliography}

\end{document}